\documentclass[twocolumn,amsmath,amssymb,aps,prx]{revtex4-2}

\usepackage{graphicx,dcolumn,bm,verbatim,hyperref,color,tabularx,enumitem,soul,floatrow,caption}
\usepackage[subrefformat=parens,labelformat=parens]{subfig}
\begin{document}
\title{A zigzag optical cavity for sensing and controlling torsional motion}
\author{Sofia Agafonova}
\author{Umang Mishra}
\author{Fritz Diorico}
\author{Onur Hosten}
 \email{onur.hosten@ist.ac.at}
\affiliation{Institute of Science and Technology Austria (ISTA), 3400 Klosterneuburg, Austria}

\begin{abstract}
    Precision sensing and manipulation of milligram-scale mechanical oscillators has attracted growing interest in the fields of table-top explorations of gravity and tests of quantum mechanics at macroscopic scales. Torsional oscillators present an opportunity in this regard due to their remarked isolation from environmental noise. For torsional motion, an effective employment of optical cavities to enhance optomechanical interactions -- as already established for linear oscillators -- so far faced certain challenges. Here, we propose a novel concept for sensing and manipulating torsional motion, where exclusively the torsional rotations of a pendulum are mapped onto the path length of a single two-mirror optical cavity. The concept inherently alleviates many limitations of previous approaches. A proof-of-principle experiment is conducted with a rigidly controlled pendulum to explore the sensing aspects of the concept and to identify practical limitations in a potential state-of-the art setup. Based on this work, we anticipate development of precision torque sensors with sensitivities below $10^{-19}~\mathrm{N\cdot m/\sqrt{Hz}}$ and with the motion of the pendulums dominated by quantum radiation pressure noise at sub-microwatts of incoming laser power. This work, therefore, paves the way to new horizons for experiments at the interface of quantum mechanics and gravity.
    
 \end{abstract}

\maketitle

\section{Introduction}
\label{sec:Introduction}

Precision sensing with mechanical oscillators has a long history in science and technology, and continues to be an active area of research in many fronts. These include detection of gravitational waves \cite{PhysRevLett.116.061102}, exploration of quantum mechanics and its potential extensions at macro scales \cite{PhysRevLett.91.130401,Pikovski2012,Chan2011,Vinante2020,Kaltenbaek2023}, tests of gravity at micro scales \cite{ARKANIHAMED1998263,PhysRevLett.124.101101,Westphal2021} or at relativistic speeds \cite{Spengler2022}, and precision determination of Newton's gravitational constant \cite{George_T_Gillies_1997} just to name a few. Recently, milligram-scale optomechanical systems \cite{PhysRevA.101.011802, PhysRevD.100.066020,Altin2017,Matsumoto2019,Catano2020} have been attracting heightened attention as promising candidates for table-top fundamental gravity and quantum mechanics experiments, since milligram-scale is expected to simultaneously allow for good optical control and large gravitational interactions \cite{Michimura2020}. This mass scale might prove beneficial for testing the ideas of gravitationally mediated entanglement \cite{Bose2017,Marletto2017,Krisnanda2020} or provide schemes for searches for dark matter targeted at a unique parameter space not covered by other potential experiments \cite{Carney_2021}. The development of sensing techniques that go beyond the state-of-the-art with milligram-scale mechanical objects is thus timely, especially in light of impressive control already achieved in more complicated systems with larger masses \cite{doi:10.1126/science.abh2634}. Our interest here is towards torsional oscillators given their natural isolation from the environmental seismic and acoustic noise, which makes them attractive for general force sensing applications. 

Numerous techniques have been developed in the framework of optomechanics to sense torsional motion via light. Optical sensing of angular displacements first flourished with optical levers, the precision versions of which date back to the Eötvös torsion balance experiments for testing the equivalence of gravitational and inertial masses \cite{von1890mathematische}. The optical lever is still the work horse of many applications, from reading-out of atomic force microscopes \cite{Alexander1989} or micro-scale torsional force sensors \cite{Pratt2023} to implementing quantum-correlation enhanced sensing protocols \cite{https://doi.org/10.48550/arxiv.2212.08197}. Nevertheless, many other techniques still continue to be developed for enhanced sensitivity, stability, or utility. These include techniques that utilize Mach-Zehnder interferometry \cite{Park:16}, Michelson interferometry \cite{doi:10.1063/5.0043098,Smetana2022}, Sagnac interferometry \cite{Martinez-Rincon:17}, or techniques that combine multi-pass optical levers with interferometry \cite{Hogan2011}. There is growing interest in optical cavity-based approaches for angular sensing, since for the related case of linear displacement sensing, these approaches have led to spectacular sensitivities \cite{PhysRevLett.116.061102} and also allowed the sought-after enhancement of quantum radiation pressure noise levels (QRPN) to above that of classical thermal noise \cite{Cripe2019}. For linear displacements, the main effect relies on the fact that mechanical displacements can change the cavity length. Probing this change with a resonant optical build-up in the cavity allows for a signal enhancement. However, for angular sensing, it is not yet clear how to most effectively utilize cavities to benefit from the same signal enhancement.

Previously investigated solutions include constructing cavities at the ends of a torsional pendulum \cite{mueller2008,mcmanus2017,PhysRevA.101.011802} and monitoring differential signals in post-processing, or constructing degenerate optical cavities for enhancing the amplitudes in higher-order cavity modes associated with beam tilting induced by small cavity mirror rotations \cite{Shimoda:22}. Both methods provide high sensitivity, and in fact the double-cavity arrangement of Ref. \cite{PhysRevA.101.011802} provides the highest torque sensitivity with milligram-scale torsional pendulums to date.  The explored approaches nevertheless have their limitations, e.g., due to far-from-optimal common mode cancellations, mechanical/thermal noise associated with complicated implementations, trade-offs between range and sensitivity, or limitations to laser power due to angular instabilities induced by the optical anti-sping effect \cite{Matsumoto2014}. Other related work for angular sensing and motion control with cavities include interfacing cavities with nanomechanical rotors \cite{Kuhn2015}.   

In this paper, we propose a novel torsion sensing concept where the yaw motion of a pendulum is mapped onto the path length of a single two-mirror optical cavity (see Fig. \ref{fig:fig1}). Further, we perform a proof-of-principle demonstration of this concept. The introduced concept inherently addresses many technical problems that can prevent precision usage of cavities for angular sensing. The cavity by design is insensitive to motion in any other mode than the yaw rotation, it does not result in a trade-off between sensitivity enhancement and usable range, it does not lead to dynamical instabilities in the form of optical anti-spring effect despite utilization of planar pendulum mirrors, and it further contains an extra intrinsic reference mode that bypasses the pendulum to directly help eliminate notorious problems with laser frequency and cavity length noises without requiring tedious absolute length or frequency stabilization.

The presented demonstration here utilizes a rigidly controlled mirror body to simulate possible motions of a real pendulum to test and validate the proposed ideas. For completeness, we also give a quantitative prediction for a system incorporating a milligram-scale torsional pendulum in an ultrahigh vacuum environment with state-of-the-art suspensions and a realistic high-finesse cavity. Based on these calculations, in the frequency range from 2 Hz to 200 Hz we anticipate a torque sensitivity of $10^{-19}-10^{-20}~\mathrm{N\cdot m/\sqrt{Hz}}$, and the domination of QRPN in the mechanical motion at sub-microwatts of incoming laser power -- much smaller than those used in previous experiments for similar physical scales. These prospects make the proposed setup interesting for macroscopic tests of quantum mechanics and tests of gravity-driven entanglement \cite{PhysRevA.101.063804,PhysRevLett.98.030405}.

The overall sensing technique is presented in Sec. \ref{sec:TheProposedSetup}, and the experimental realization is described in Sec. \ref{sec:ExclusiveTorsionSensitivity}. The characterization of the coupling of different motions to the cavity, as well as a simple demonstration of the utility of the reference mode for mitigating noise is presented in Sec. \ref{sec:LaserFrequencyStabilization}. Finally, a realistic noise model to predict the performance of a future milligram-scale torsional pendulum integrating the developed concept is presented in Sec. \ref{sec:ExpectedSensitivity}.

\section{The sensing principle}
\label{sec:TheProposedSetup}

\begin{figure}
    \includegraphics{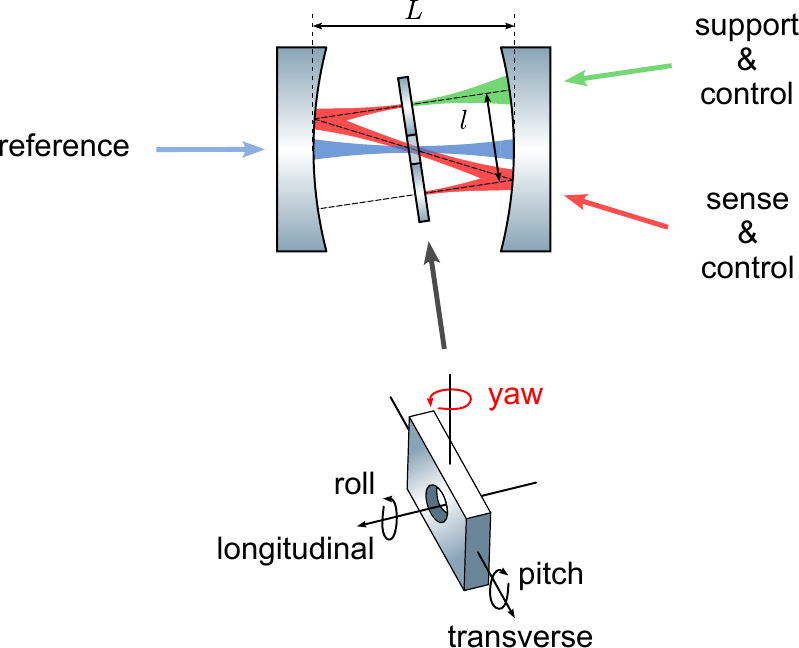}
    \caption{\label{fig:fig1} Schematics of the proposed torsion sensing concept. A suspended torsional pendulum with a hole in the center and high reflectivity coatings on both faces is inserted into a two-mirror optical cavity. The system operates around a specific yaw orientation that fixes the zigzag beam separation $l$. See text for additional details. } 
\end{figure}

The proposed setup is an optical cavity formed by two identical spherical mirrors and a torsional pendulum with high reflectivity coatings on both sides (Fig. \ref{fig:fig1}). The pendulum has a hole in the center to allow for coupling to the cavity mode highlighted in red, which we call the zigzag mode. This mode is exclusively sensitive to the yaw motion, whereby the mode path length -- hence the resonance frequency -- is linearly proportional to the yaw rotation angle around the central operating angle of the pendulum (Fig. \ref{fig:fig1}). If realized ideally, linear displacements (transverse or longitudinal) or roll motion of the pendulum have no effect on the zigzag mode path length, leaving the resonance frequency unaltered. Furthermore, the pitch motion can be made to induce path length changes that are only quadratic in the pitch angle due to the symmetry of the geometry. The zigzag mode is thus ideally suited to sensing the yaw motion or manipulating it by means of radiation pressure, e.g., in the form of optical spring or feedback damping \cite{PhysRevLett.99.160801}. Note that, as we will illustrate, sensing the yaw motion simply amounts to optically tracking the cavity resonance frequency. In addition to the zigzag mode, the hole in the pendulum also allows for a separate cavity mode highlighted in blue, which we call the on-axis mode. This cavity mode is decoupled from any motion of the pendulum, and serves as a relative length reference for the zigzag mode. Offset locking the frequency of the sensing laser to this mode mitigates laser frequency noise and also eliminates the effects of cavity length drifts from the measurements carried out on the zigzag mode. Lastly, the cavity mode highlighted in green, which we call the support mode, allows for application of an additional radiation pressure torque on the pendulum (e.g., if the restoring torque supplied by the pendulum suspension is too weak), or for monitoring and controlling the longitudinal linear motion of the pendulum.

\begin{figure*}[t]
    \sidesubfloat[]{\label{fig:fig2_1}
    \includegraphics[]{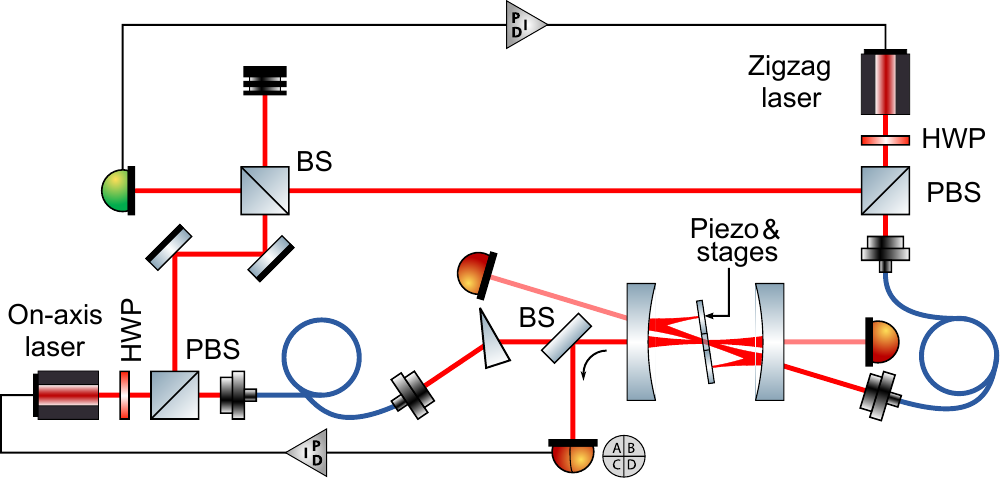}}\hfill
    \sidesubfloat[]{\label{fig:fig2_2}\includegraphics[]{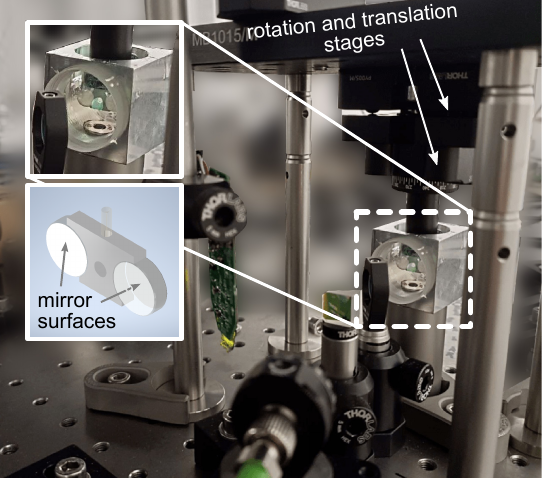}}
    
    \vspace{5mm}
    \sidesubfloat[t]{\label{fig:fig2_3}
    \includegraphics[]{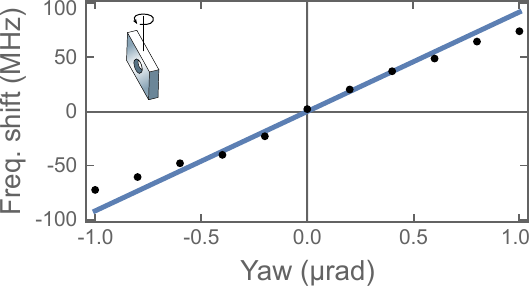}}   \hfill
    \sidesubfloat[t]{\label{fig:fig2_4}
    \includegraphics[]{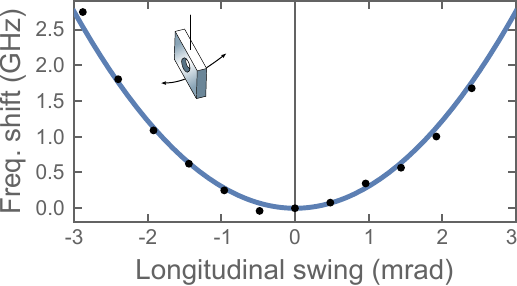}}  \hfill
    \sidesubfloat[t]{\label{fig:fig2_5}
    \includegraphics[]{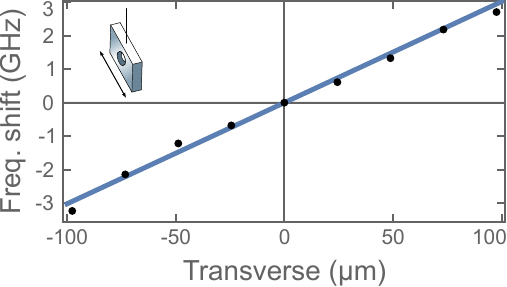}}
    
    \caption{\label{fig:fig2}(a) Experimental setup. The 'on-axis' and the 'zigzag' lasers are coupled to their respective cavity modes. The on-axis laser frequency is locked to the cavity. A beatnote between the lasers obtained on a fast photodetector is used to lock the zigzag laser frequency to the on-axis laser with a variable frequency offset. The position of the pendulum is manipulated via a rotation stage, a 5-axis stage (Thorlabs PY005/M), and an attached piezo transducer. BS: beam splitter, PBS: polarizing beam splitter, HWP: half-wave plate. (b) A picture of the cavity. (c-e) Measured zigzag mode frequency shifts, as a function of the yaw (c), longitudinal swing (d), and transverse (e) displacements of the pendulum. Solid lines represent parameter-free models in (c) and (d), and a linear fit to the data in (e). Data was acquired via discrete changes in the stages in (d) and (e), and via continuous piezo scans in (c).}
\end{figure*}

The angular sensitivity provided by the technique can be quantified by first noting the zigzag mode frequency shift per yaw rotation: $\frac{c}{\lambda}\frac{l}{L_{zig}}$ (Hz/rad). Here $\lambda$ is the wavelength of the utilized light, $l$ is the zigzag beam separation, $L_{zig}\approx2L-\frac{l^2}{2L}\frac{g}{1+g}\approx2L$ is the geometrical length of the zigzag mode, $L$ is the separation between the two spherical mirrors, and $g=1-\frac{L}{R}$ is the commonly used cavity stability parameter. A metrologically relevant sensitivity parameter is the ratio of the induced shift to the full linewidth of the zigzag mode. This can be expressed using the cavity finesse $\mathcal{F}$ (ratio of the mode's free spectral range to its linewidth) as
\begin{equation}
\label{eq:sensitivity}
    S=\frac{2l}{\lambda}\mathcal{F} \quad (\mathrm{rad}^{-1}).
\end{equation}
Note that $S$ is independent of the cavity length, and with this definition, $n\sim1/(S\ \delta\theta)^2$ has the interpretation of the total number of photons required to achieve an angular resolution of $\delta\theta$ -- assuming $n\gg 1$.

In the outlined cavity geometry, the zigzag mode is a stable cavity mode in the Gaussian-beam propagation sense as long as its length $L_{zig}$ is equal to or smaller than twice the radius of curvature $R$ of the spherical mirrors. To a good approximation, this corresponds to $L\leq R$. In terms of the cavity stability parameter, this condition is given by $g\geq 0$. Assuming there exists a static equilibrium angle for the pendulum, the optomechanical system does not possess an angular instability (again for $g\geq 0$) originating from the geometric anti-spring effect discussed in Ref. \cite{Matsumoto2014}. Instead, one obtains a spring effect that increases the restoring torques. Although we will not discuss this effect further, we just note this physically means that in response to a rotation, the zigzag mode location at the pendulum shifts such that it acts to restore the orientation by changing the lever arm of the radiation pressure force -- and hence the applied torque. The same arguments are also valid for any motion that involves the pitch rotation (Fig. \ref{fig:fig1}) of the pendulum bar.

Since we are utilizing a rigidly controlled mirror body to simulate a pendulum in the current demonstration, the discussed effects of radiation pressure will not yet be relevant. For the same reasons, the support mode will not be discussed further.

\section{Experimental realization}
\label{sec:ExclusiveTorsionSensitivity}

Our experimental setup is shown in Fig. \subref*{fig:fig2_1}. Two 780 nm DFB lasers are used to couple to the zigzag and the on-axis modes of the cavity -- typically a few hundred microwatts of power incident for each. The frequency of the `on-axis laser' is locked to the cavity with the `squash locking' method \cite{https://doi.org/10.48550/arxiv.2203.04550} developed in our lab, which here utilizes a slight ellipticity in the beam shape provided by a prism, and a quadrant photodetector for its measurement. The zigzag laser is locked to the on-axis laser with a variable GHz-range frequency offset by means of stabilizing the beat note between the two lasers obtained on a fast photodetector \cite{PhysRevApplied.17.054031}. 

A picture of the zigzag cavity is shown in Fig.\subref*{fig:fig2_2}. The simulated pendulum is composed of two flat mirrors (2 mm thick, 7 mm in diameter) that are fixed on an aluminum holder with a center-to-center separation of 11 mm (see insets in Fig. \subref*{fig:fig2_2}). The holder hosts a 2 mm diameter hole in its center. This assembly is suspended from above with a rigid post that is mounted on a 5-axis tip-tilt-translation stage in addition to a rotation stage. The effective lever arm length for tilting the pendulum (to simulate swing motion) is 6 cm. For the cavity, we use two spherical mirrors with $R=50$ mm radius of curvature epoxied on an aluminum cavity spacer. 

In this work, we employ a special strategy to construct the cavity to ease the optical alignment to the zigzag mode. This strategy requires the choice of a specific cavity length $L$ such that, in absence of the pendulum, the cavity hosts the mode depicted in dotted lines in Fig. \ref{fig:fig1} -- the no-pendulum zigzag mode. Spatially, the ends of this mode are perpendicular to the spherical mirrors and also perpendicular to the pendulum once it is inserted in the cavity. In Appendix \ref{sec:NoPendulumZigZag} we show that geometrically such a mode can exist in a cavity if $g>1/2$. Under this condition, the separation of the parallel no-pendulum zigzag beams $l$ (Fig. \ref{fig:fig1}) is given by 
\begin{equation}
\label{eq:l}
    l=R\sqrt{1-\frac{1}{4}\frac{1}{g^2+2g-1}}.
\end{equation}
Matching this separation to the center-to-center mirror separation of the pendulum (i.e., $l=11$ mm) requires $L$ to be 24.8 mm (equivalently, $g=0.504$) for $R=50$ mm. Before gluing the mirrors to the cavity spacer, we ensure that we indeed realize this particular cavity length by coupling light to the on-axis cavity mode and monitoring the transverse mode separation which directly reveals the cavity length given a known $R$ \cite{Siegman1986}. For our parameters, this procedure amounts to setting the separation between the fundamental and third order transverse cavity modes to 28.3 MHz. Once the cavity is made, the light is first coupled to the no-pendulum zigzag mode, then the pendulum is inserted and the yaw angle of the pendulum is tweaked to recover transmission from the cavity to conclude the alignment.

The resulting cavity here has a free spectral range of 6.2 GHz for the on-axis mode and 3.2 GHz for the zigzag mode. The full linewidths for the two modes are 7 MHz and 14 MHz respectively. The finesses associated with the on-axis and zigzag modes are then $\mathcal{F}_{on}=880$ and $\mathcal{F}_{zig}=230$, with the difference in the finesse practically accounted for by the tripling of the number of roundtrip bounces (hence the light loss from the cavity). The central operating yaw angle for the pendulum is 8.5 degrees from the cavity axis.

We would like to clarify that although the no-pendulum zigzag mode here exists in a geometric optics sense, it is strictly speaking unstable from a Gaussian beam propagation perspective. Nevertheless it still shows power buildup at resonance conditions, and furthermore, due to its unstable nature, resonances can be observed for a larger range of input beam incident angles, making it easier to find a starting point for the alignment. We also note that the procedure described here for constructing and aligning the cavity is just a matter of choice. In particular, a specific cavity length is not required to have a zigzag mode (in presence of the pendulum): for a given $L$ there always exists a pendulum yaw orientation that results into a desired $l$ as long as $g\geq 0$.  

\section{Results}
\label{sec:LaserFrequencyStabilization}
To characterize the sensitivity of the setup to different motions of the simulated pendulum, we track the zigzag mode resonance frequency as we manipulate the position and orientation of the pendulum using the 5-axis stage. Frequency tracking is accomplished by means of scanning the zigzag laser frequency while monitoring the transmission, and the pendulum manipulation is accomplished either with manual actuation or with an attached piezo actuator slab. We compare the observed frequency shifts to the expected ones for a specific pendulum movement by calculating the expected zigzag mode length changes $\delta s$ as outlined in Appendix \ref{sec:ZigZagModePathCalculation}. These length changes result in a cavity frequency shift $\delta \nu=\frac{c}{\lambda}\frac{1}{s}\delta s$, where $s$ is the zigzag mode length itself. The results are presented in Fig. \subref*{fig:fig2_3}, \subref*{fig:fig2_4} and \subref*{fig:fig2_5}. 

First, we confirm the $\sim$85 MHz/$\mathrm{\mu}$rad linear dependence ($S\approx6.5\times10^6$ $\mathrm{rad}^{-1}$) of the resonance frequency on the pendulum's yaw angle (Fig. \subref*{fig:fig2_3}). The deviation of the data from linear behavior near $\pm 1 \mu$rad is an artifact rooted in the piezo actuation nonlinearities. With manual mechanical actuation, we have verified that the linear behavior persists even at milliradian levels. For future optomechanics experiments, the relevant angular variations will not be larger than the microradian level.

Next, we observe the predicted quadratic change in the resonance frequency in response to tilting the pendulum assembly to mimic the longitudinal swing motion (Fig. \subref*{fig:fig2_4}). For the pendulum body, this motion is a combination of a pure pitch rotation about its centroid and a pure longitudinal translation. Here, the observed resonance shifts originate solely from the pitch angular rotation component of the motion -- giving rise to a $\frac{c}{\lambda}\frac{1+g}{4g}$ ($\mathrm{Hz/rad}^2$) cavity shift per squared pitch angle (a second order sensitivity of $S^{(2)}=\frac{L}{\lambda}\frac{1+g}{g}\mathcal{F}~~(\mathrm{rad}^{-2})$) -- with no measurable contribution from the longitudinal translation (see Appendix \ref{sec:OnRollLongSens}). These findings show that there is a specific alignment of the pendulum and the cavity -- the bottom of the parabola -- which makes the system almost blind to the pitch motion of the pendulum.

Lastly, we observe an initially-unconsidered effect, which manifests in the form of a residual sensitivity to transverse translations (Fig. \subref*{fig:fig2_5}). For a realistic estimation of achievable yaw sensitivities, this effect needs to be taken into account to assess signal leakages from the transverse swing and the roll motions of the pendulum. As we will elaborate later with an informative comparison, the involved cavity shifts are nevertheless subdominant. Appendix \ref{sec:OnRollLongSens} contains an analysis of the effect together with more data. The analysis indicates that the response originates from a relative tilt $\delta\alpha$ of the two mirrors (about the yaw axis) that form the pendulum which are imperfectly glued to the aluminum holder. The model predicts a $\frac{c}{\lambda}\frac{\delta\alpha}{2L}$ ($\mathrm{Hz/m}$) cavity frequency shift per transverse pendulum translation -- equivalent to a sensitivity of $S=\frac{2\delta\alpha}{\lambda}\mathcal{F}~~(\mathrm{m}^{-1})$. With the observed $30~\mathrm{MHz}/\mu\mathrm{m}$ response, this leads to an inferred relative tilt of $\delta\alpha=0.2$ degrees between the two mirrors of the constructed pendulum. An initial comparison of this residual sensitivity with the main yaw-rotation sensitivity can be made by converting the later into a sensitivity to pendulum-end-point displacements. This comparison indicates a $2/\delta\alpha=570$ times stronger sensitivity for the yaw motion.

\begin{figure}[t]
  \includegraphics[]{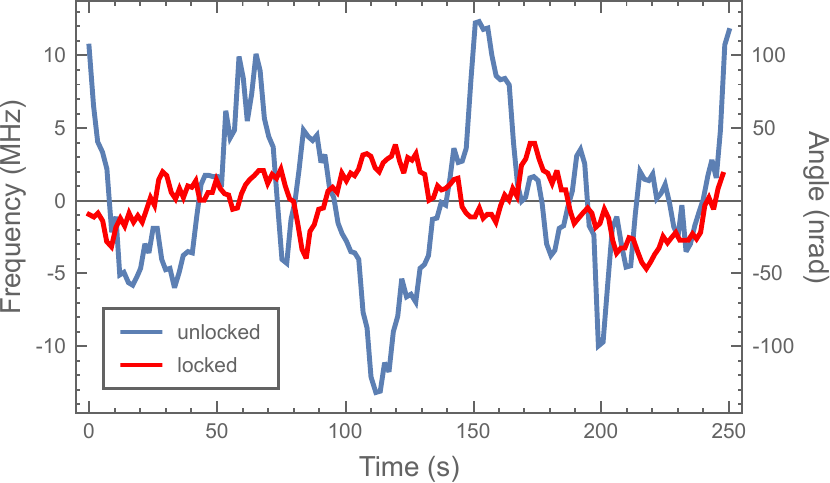}
  \caption{\label{fig:fig3} Relative frequency drifts between the zigzag cavity mode and the zigzag laser frequency with and without offset locking the zigzag laser to the on-axis mode.}
\end{figure}

Having characterized the response to various motions, we return to an important technical advantage brought by the design, namely the existence of a reference mode that is decoupled from the pendulum. Referencing the zigzag laser frequency to the on-axis mode of the cavity -- as already discussed -- allows for automatic elimination of the cavity length noise and for mitigation of laser frequency noise. To demonstrate the utility of the reference mode, we compare the relative frequency between the zigzag laser and the zigzag cavity mode as a function of time, both when the zigzag laser is free-running and when it is offset locked. The results are shown in Fig. \ref{fig:fig3}. In these measurements, the relative laser-cavity frequencies were identified by scanning over the resonance by modulating the yaw motion for a fixed laser offset locking point. As qualitatively evidenced, engaging the frequency locking reduces the overall noise (in this case $\sim$3 fold). It is important to note that in the current setup the fluctuations in the on-axis laser lock and the offset lock themselves are at least an order of magnitude smaller than the observed fluctuations for the red curve in Fig. \ref{fig:fig3}. This indicates that these measurements truly reflect some mechanical drifts in the yaw motion of the simulated pendulum -- free from cavity length and laser frequency drifts. Carrying out more quantitative  spectral measurements for noise immunity is not very meaningful at this stage until an upgraded compact system that is housed in a vacuum environment is constructed.

\section{Discussion and Outlook}
\label{sec:ExpectedSensitivity}

Here we proposed and demonstrated a novel cavity-based angle sensing concept which responds sensitively only to yaw motion of a suspended pendulum. We would like to put the system and its first characterizations into context by discussing the system's intended application to sensing with milligram-scale torsional pendulums.

First, we would like to point to an advantageous property of the demonstrated scheme: Although the sensitivity increases with cavity finesse (Eq. \ref{eq:sensitivity}), the accessible angular measurement range remains the same -- it is determined purely by geometrical factors. Such a property does not hold in general for other explored schemes, as exemplified in the case of ref. \cite{Shimoda:22}, where the angular sensitivity increases with cavity finesse at the expense of a drop in the sensing range. In our setup, the limitation comes form the fact that the geometry of the zigzag mode slowly walks off from the initial mode geometry as the yaw angle is scanned -- until the input light is no longer coupled to the zigzag mode. With this effect, the sensing range is limited to $\theta_{rng} = \frac{2g\lambda}{\pi \mathrm{w}_0}$ (see Appendix \ref{sec:range}), where $\mathrm{w}_0$ is the beam waist size at the cavity center. In the current setup, $\mathrm{w}_0\approx75\mu$m, leading to $\theta_{rng}\approx 0.2$ degrees. Note that, throughout the angle scan, the laser frequency can be easily made to track the cavity resonance, making the finesse-dependent cavity linewidth not a limitation to the sensing range.

We now discuss the unexpected residual sensitivity of the cavity to the transverse motion originating from the non-parallel nature of the mirrors at the two ends of the pendulum. Note that, for a monolithic pendulum mirror, one might think that this effect could completely disappear. However, low-loss optical coatings required for high-finesse cavities invariably lead to material stress, resulting in some residual curvature especially for thin (e.g., $\sim$500 $\mu$m) substrates. Nevertheless, realistically, the relative tilt of the two ends could be at least an order of magnitude smaller than the 0.2 degrees inferred in the current work -- assuming several meters of radius-of-curvature.

To operationally compare the residual transverse motion sensitivity (Fig. \subref*{fig:fig2_3}) to the main yaw motion sensitivity (Fig. \subref*{fig:fig2_5}), we first compare the cavity shifts caused by the rms motion of individual normal modes of a pendulum that is assumed to be thermal noise limited in its motion. We will than substantiate this discussion with more a informative comparison of spectral noise distributions. We assume a $12\times0.5\times0.5$ mm\textsuperscript{3} pendulum suspended on a 5 cm long 1 $\mu$m thick fiber in the experimentally studied zigzag cavity, and assign $k_{B}T/2$ worth of energy to transverse-swing, roll, and yaw modes. Note that pitch, longitudinal-swing and -violin modes will couple to the measurement quadratically and will thus have negligible contributions. Assuming specific frequencies for the considered modes (2 Hz swing, 2 Hz roll, and 5 mHz yaw), the rms fluctuations in the corresponding coordinates could be calculated and converted to zigzag mode frequency shifts using the measured sensitivities. Carrying out this calculation also requires an understanding of how each pendulum mode couples to the zigzag cavity mode, which is detailed in Appendix \ref{sec:OnRollLongSens}. Following this procedure, one obtains rms cavity shifts of about 75 kHz and 4 kHz respectively for transverse-swing and roll motions, while obtaining 25 GHz for yaw motion. This indicates a vast dominance of the signal by the yaw motion.

\begin{figure}[t]
  \includegraphics{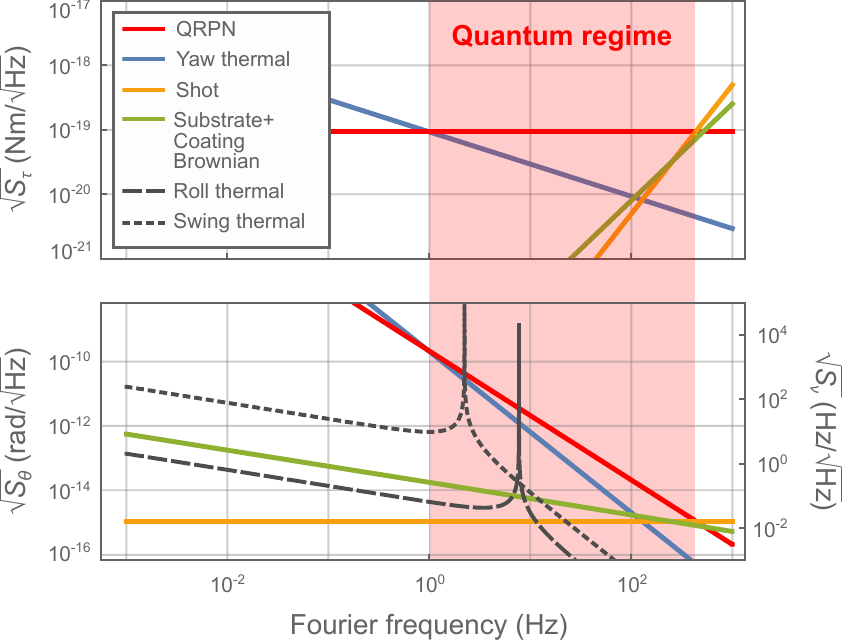}
  \caption{\label{fig:fig4} Individual contributions of noise sources in an envisioned future milligram-scale torsional pendulum employing a zigzag cavity. Expected angular noise density $\sqrt{S_{\theta}}$ and the equivalent torque noise density $\sqrt{S_{\tau}}$ are plotted in accordance with the discussions in the text for system parameters given in Table \ref{tab:my_label}. The right-hand-side axis of the bottom plot indicates the corresponding cavity frequency noise density $\sqrt{S_{\nu}}$ that would actually be measured for the considered system.}
\end{figure}

Finally, we present a spectral estimation of the performance that could be expected from a future milligram-scale torsional pendulum that utilizes a zigzag cavity. Note that, certain system parameters assumed from this point on differ from those utilized so far. Assuming other potential technical noises are under control, we consider four inevitable noise sources: suspension thermal noise $S_{\tau}^{th}$ \cite{PhysRevA.101.011802}, quantum radiation pressure noise $S_{\tau}^{QRPN}$ \cite{Cripe2019}, photon shot noise $S_{\theta}^{shot}$ \cite{Michimura2020}, and mirror substrate/coating Brownian noise $S_{\theta}^{th}$ \cite{PhysRevLett.91.260602}. The respective power spectral densities for these noise contributions are given by

\begin{equation} \label{eq:noise}
    \begin{split}
    &S_{\tau}^{th}(\omega) = 4k_{B}TI\gamma_{m},\\
    &S_{\tau}^{QRPN}(\omega) = \frac{8l^{2}\mathcal{F}^{2}\hbar\omega_{L}P_{in}}{\pi^{2}c^{2}},\\
    &S_{\theta}^{shot}(\omega) = \frac{\hbar}{S_{\tau}^{QRPN}},\\
    &S_{\theta}^{th}(\omega) = \tfrac{16k_{B}T}{\omega l^{2}}\tfrac{1-\sigma^{2}}{\sqrt{\pi}E\mathrm{w}_{0}}\left(\phi_{sub}+\tfrac{2}{\sqrt{\pi}}\tfrac{1-2\sigma}{1-\sigma}\tfrac{d}{\mathrm{w}_{0}}\phi_{coat}\right).
    \end{split}
\end{equation}

Here, $\omega=2\pi f$ is the angular frequency -- with $f$ being the Fourier frequency used in Fig.\ref{fig:fig4}. The subscripts $\theta$ and $\tau$ refer to angle and torque variables, and their associated spectral densities can be converted into each other through the frequency dependent susceptibility $\chi(\omega)$ of the torsional oscillator \cite{Michimura2020}: $S_{\theta}(\omega)=|\chi(\omega)|^2 S_{\tau}(\omega)$. In Eq. \ref{eq:noise},  $k_{B}$ is the Boltzmann constant, $\hbar$ is the reduced Planck constant, and $c$ is the speed of light. The remaining parameter definitions together with their assumed values are listed in Table \ref{tab:my_label}.

\begin{table}[t]
    \centering
    \begin{tabular}{|c|c|c|}
        \hline
       parameter & description & value\\
       \hline
       $T$ & temperature & 300 K \\
       $l$ & pendulum width & 5 mm\\
       $h$ & pendulum height & 1 mm\\
       $t$ & pendulum thickness & 0.5 mm\\
       $m$ & pendulum mass & 6 mg \\
       $I$ & pendulum moment of inertia & $ml^{2}/12$\\
       $D$  & fused silica suspension fiber diameter & 1 $\mu$m\\
       $\omega_{m}/2\pi$ & mechanical (torsional) frequency & 10 mHz\\
       $Q_{m}$ & yaw mode quality factor \cite{gretarsson1999} & $2\times 10^{4}$\\
       $\gamma_{m}$ & yaw mode damping rate & $\frac{\omega_{m}}{Q_{m}}\frac{\omega_{m}}{\omega}$\\
       $\sigma$ & mirrors' substrate Poisson's ratio \cite{Gregory_M_Harry_2002} & 0.15\\
       $E$ & mirrors' substrate Young's modulus \cite{Gregory_M_Harry_2002} & 70 GPa\\
       $\mathrm{w}_{0}$ & beam radii at the mirrors & 100 $\mu$m\\
       $\phi_{sub}$ & mirrors' substrate loss angle \cite{,PhysRevLett.91.260602, RevModPhys.86.121} & $10^{-7}$\\
       $\phi_{coat}$ & mirrors' coating loss angle \cite{PhysRevLett.91.260602, Gregory_M_Harry_2002} & $10^{-4}$\\
       $d$ & coating's thickness & 10 $\mu$m\\
       $\mathcal{F}$ & cavity finesse \cite{hosten2016} & $2\times 10^{4}$\\
       $\lambda_{L}$ & sensing laser wavelength & 780 nm\\
       $\omega_{L}$ & sensing laser frequency & $2\pi c/\lambda_{L}$\\
       $P_{in}$ & incident laser power & 0.4 $\mu$W\\
       \hline
    \end{tabular}
    \caption{\label{tab:my_label} Anticipated experimental parameters of the zigzag setup including a mg-scale torsion pendulum.}
\end{table}

Assuming a 5-cm long cavity with 10-cm radius-of-curvature mirrors, the resulting torque and angle noise contributions that are expected to limit the yaw sensitivity are presented in Fig. \ref{fig:fig4}. In addition to the sources of noise already discussed in Eq. \ref{eq:noise}, here we have also included limitations due to leakage signals from the transverse-swing and roll modes of the pendulum (dotted and dashed lines). The motion of these modes are assumed to be thermal-noise limited, and their gravitational-dissipation-dilution-enhanced \cite{Michimura2020} mechanical quality factors are assumed to be $10^6$. A 60-mdeg residual relative yaw-bending of the two ends of the pendulum is assumed for the cause of the leakage signals, whose origins are detailed in Appendix \ref{sec:OnRollLongSens}. 

Based on the presented calculations, in the $\sim2-200$ Hz frequency range, the system could be expected to be limited by QRPN with only sub-microwatts of input optical power -- meaning that the mechanical motion will be dominated by quantum noise. This renders the system interesting for experimental tests of macroscopic quantum mechanics or gravity-driven entanglement \cite{PhysRevA.101.063804,PhysRevLett.98.030405}. Furthermore, a thermal noise limited torque sensitivity of $10^{-19}-10^{-20}~\mathrm{N\cdot m/\sqrt{Hz}}$ is anticipated in this frequency range -- assuming input optical power is adjusted as needed to lower the QRPN. As a reference, the current record in the milligram scale is $2\times10^{-17}~\mathrm{N\cdot m/\sqrt{Hz}}$ \cite{PhysRevA.101.011802}, highlighting the potential of the envisioned system as a torque sensor.

\section*{Acknowledgements}
We thank Pere Rosselló for his contributions to the initial modeling of the presented sensing technique. This work was supported by Institute of Science and Technology Austria, and the European Research Council under Grant No. 101087907 (ERC
CoG QuHAMP).

\section*{Appendix}

\appendix

\begin{figure*}[t!]
  \includegraphics[]{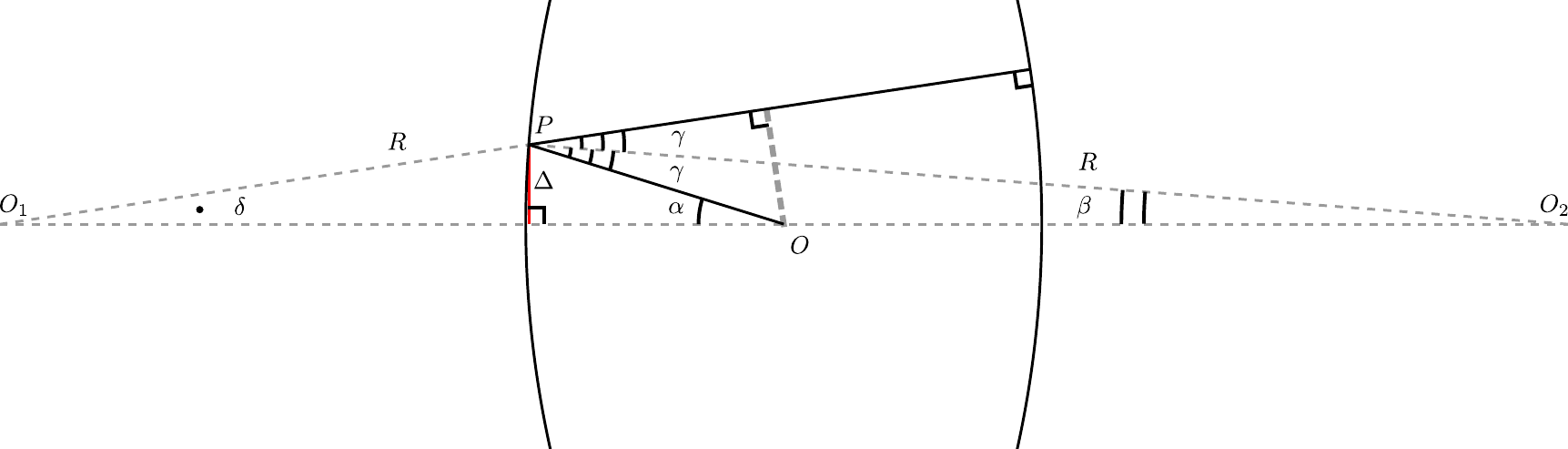}
  \caption{\label{fig:figA1} Illustration for Appendix \ref{sec:NoPendulumZigZag}.}
\end{figure*}

\section{No-pendulum zigzag mode}
\label{sec:NoPendulumZigZag}

In this appendix we show that the zigzag mode in the experimentally utilized cavity also exists in absence of the pendulum as a geometric self-replicating path, and we relate the cavity configuration to the zigzag beam separation. From the geometric point of view, showing the existence of the mode in the cavity is equivalent to finding an angle $\alpha$ (Fig. \ref{fig:figA1}) at which one can send a beam of light from the center of the cavity to one of the spherical mirrors such that the beam reflected from this mirror is perpendicular to the other spherical mirror at the point of contact. We can write

\begin{equation}
    \begin{cases}
    R\sin\beta=(\frac{L}{2}-R(1-\cos\beta))\tan\alpha\\
    R\sin\beta=(R-L+R(1-\cos\beta))\tan\delta\\
    \alpha+\delta=2\gamma\\
    \beta+\gamma=\alpha.
    \end{cases}
\end{equation}
The first equation expresses $\Delta$ through angles $\beta$ and $\alpha$, the second one — through $\beta$ and $\delta$, and the last two equations consider triangles $OO_{1}P$ and $OO_{2}P$ respectively. Switching to new variables $x=\tan\alpha,~y=\tan\beta,~z=\tan\gamma,~t=\tan\delta$ and expressing $L/R$ through $1-g$, we can rewrite the same equations as

\begin{equation}
\begin{cases}
    y=(1-\frac{g+1}{2}\sqrt{1+y^2})x\\
    y=(\sqrt{1+y^2}(g+1)-1)t\\
    \frac{x+t}{1-xt}=\frac{2z}{1-z^2}\\
    \frac{y+z}{1-yz}=x.
\end{cases}
\end{equation}

From these equations, one can first express $x$, $y$ and $t$ as a function of $z$ as $x=\frac{4z}{3-z^2},~y=\frac{z}{3},~t=\frac{2z}{3+z^2}$. Then, one can also express $z$ through $g$, eliminating all other variables to obtain  

\begin{equation}
    t=\tan\delta=\frac{\sqrt{(2g+5)(2g-1)}}{2g^2+4g-1},
\end{equation}
which leads us to the conclusion that the sought-after configuration only exists if $g>1/2$, or if $L<R/2$. The separation between the parallel no-pendulum zigzag beams is then given by $l=2(R-L/2)\sin\delta=R\sin\delta(1+g)$, which gives Eq. \ref{eq:l}.

\section{Zigzag mode path calculation}
\label{sec:ZigZagModePathCalculation}

\begin{figure*}[]
    \centering
    \includegraphics{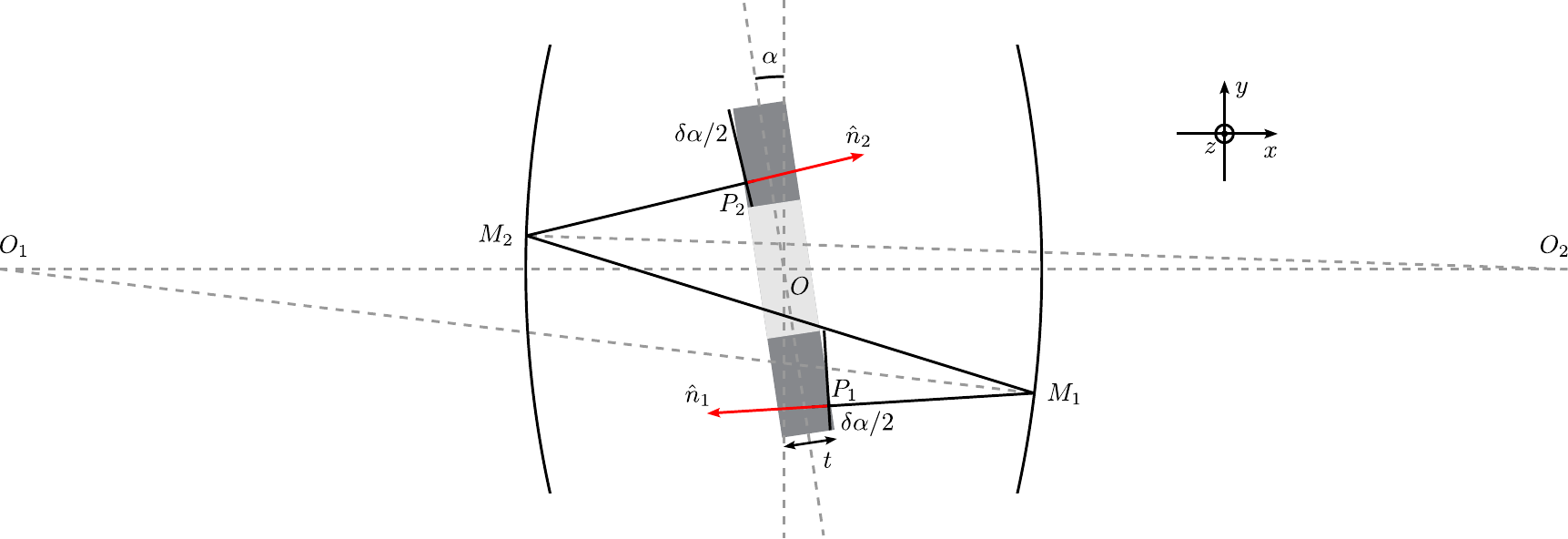}
    \caption{\label{fig:figA2} Illustration for Appendix \ref{sec:ZigZagModePathCalculation}.}
\end{figure*}

In this appendix, we describe our procedure for calculating the zigzag mode round trip length as a function of the pendulum's location and orientation. The calculations take into account the possible angular misalignment of the two ends of the pendulum. This procedure forms the basis for a general numerical computation of the path lengths, and it also allows for analytical approximations in limiting cases, which are readily utilized in the main text for evaluating cavity frequency shifts. 

First, we concretely define the geometrical problem, and introduce the points $M_{1}$ and $M_{2}$, on the right- and left-sides of the origin $O$, at which the zigzag mode reflects from the cavity mirrors (Fig. \ref{fig:figA2}). These points are constrained to the surfaces of the spherical mirrors, and are parametrized by the cavity length $L$ and the mirrors' radius-of-curvature $R$. We then introduce the points $P_1$ and $P_2$ where the reflections from the pendulum mirror surfaces take place. These points are constrained to the surfaces of the respective pendulum mirrors, and are parametrized by the orientation, location, and thickness of the pendulum. The closing of the zigzag path onto itself requires the rays $M_1P_1$ and $M_2P_2$ to be at normal incidence to the two pendulum mirror surfaces, i.e., parallel to the surface normal unit vectors $\hat{n}_i=\mp\{\cos{\alpha_i\cos{\beta_i},\sin{\alpha_i}\cos{\beta_i},\sin{\beta_i}}\}$, with $i=\{1,2\}$. Here $\alpha_i$ and $\beta_i$ refer to the yaw and pitch angles respectively. $\hat{n}_{1,2}$ might differ in their orientation, for example in their yaw angle $\alpha_{1,2}=\alpha\mp\delta\alpha/2$ as illustrated in Fig. \ref{fig:figA2}.   

 We proceed by solving for the points $M_1$ and $M_2$ by requiring the spherical mirror normal vectors $\overrightarrow{M_2O_2}$ and $\overrightarrow{M_1O_1}$ to bisect the reflecting rays. This is mathematically expressed as 
\begin{equation}
\label{eq:angle_condition}
    \begin{cases}
        \frac{\overrightarrow{M_{1}M_{2}}}{|M_{1}M_{2}|}+\hat{n}_{1}=\lambda_{1}~\overrightarrow{M_{1}O_{1}}\\[0.5em]
        \frac{\overrightarrow{M_{2}M_{1}}}{\left|M_{2}M_{1}\right|}+\hat{n}_{2}=\lambda_{2}~\overrightarrow{M_{2}O_{2}}.
    \end{cases}
\end{equation}

 Here $\lambda_{1,2}$ are unknown proportionality constants. Including the constraints on $M_{1,2}$, the system solves for eight variables (three for each of $M_{1}$ and $M_{2}$, and two for real numbers $\lambda_{1,2}$) and contains eight equations, and is therefore complete. The inferred points are functions of cavity parameters $L$ and $R$, the pendulum's yaw and pitch angles and the relative bending of the pendulum ends.

 Next, we solve for the points $P_{1,2}$ utilizing the fact that they are constrained to the pendulum mirror surfaces. Given the already found $M_{1,2}$ and the known directions $\hat{n}_{1,2}$ of the rays towards the pendulum, this amounts to finding the intersection of the rays with the surfaces. The latter are parametrized with thickness $t$ of the pendulum and its location in $xy$-plane in addition to the normal vectors.

Following the outlined steps one can calculate the length of $P_{1}M_{1}M_{2}P_{2}$ to obtain the zigzag mode length, and to evaluate its resonance frequency changes in response to the pendulum motion. Note that there is also a contribution to cavity resonance shifts due to changes in the Gouy phase shifts \cite{nagourney2010}, but it is omitted since changes of geometric origin always dominate for any real pendulum motion in the investigated configuration.

\section{Translation and roll sensitivities}
\label{sec:OnRollLongSens}
\begin{figure}[t]
    \centering
    \includegraphics{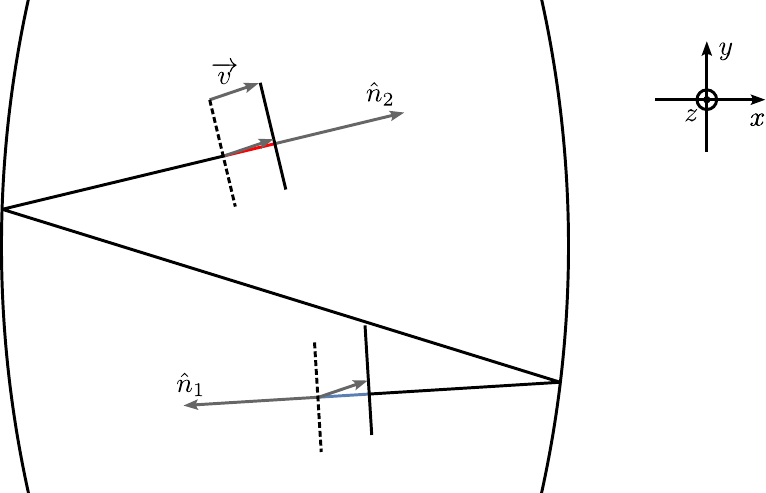}
    \caption{\label{fig:figA3} Illustration for Appendix \ref{sec:OnRollLongSens}.}
\end{figure}

In this appendix, we will focus on the effects caused by arbitrary translations and roll rotations of the pendulum. Note that in this section $\alpha$, $\beta$, and $\gamma$ stand for yaw, pitch, and roll angles of the pendulum respectively with no connection to definitions in Appendix \ref{sec:NoPendulumZigZag}. We will start with the effects on the cavity due to the translations. When the pendulum is translated by a vector $\vec{v}$ (Fig. \ref{fig:figA3}), the zigzag mode path length changes by $\delta s$ based on the projections of $\vec{v}$ onto the pendulum mirror normal vectors:
\begin{equation}
    \label{eq:projection}
    \delta s=\hat{n}_1\cdot\vec{v}+\hat{n}_2\cdot\vec{v}
\end{equation}
The unit vectors $\hat{n}_{1,2}$ encode the information on both the central angles and the misalignment angles for yaw $\alpha_{1,2}=\alpha\mp\delta\alpha/2$  and pitch $\beta_{1,2}=\beta\mp\delta\beta/2$ degrees of freedom. To lowest order in $\delta\alpha$ and $\delta\beta$ the unit vectors can be expressed as
\begin{equation}
\label{eq:unitvecs}
    \begin{split}
        &\hat{n}_{1,2}=\mp \hat{n}^{(0)}+\cos\beta~\frac{\delta\alpha}{2}~\hat{n}^{(\delta\alpha)}+\frac{\delta\beta}{2}~\hat{n}^{(\delta\beta)}\\
        &\hat{n}^{(0)}=\{\cos\alpha \cos\beta,\sin\alpha \cos\beta,\sin\beta\}\\
        &\hat{n}^{(\delta\alpha)}=\{-\sin\alpha, \cos\alpha,0\}\\
        &\hat{n}^{(\delta\beta)}=\{-\cos\alpha \sin\beta,-\sin\alpha \sin\beta,\cos\beta\}.
    \end{split}
\end{equation}
Here $\hat{n}^{(0)}$, $\hat{n}^{(\delta\alpha)}$ and $\hat{n}^{(\delta\beta)}$ are mutually orthogonal unit vectors that span a Cartesian basis. Assuming a central pitch angle $\beta\ll1$, such that $\cos\beta\approx1$, the path length change resulting from a pendulum translation can be expressed simply as
\begin{equation}
\label{eq:pathchange}
    \delta s\approx(\vec{v}\cdot\hat{n}^{(\delta\alpha)})~\delta\alpha+(\vec{v}\cdot\hat{n}^{(\delta\beta)})~\delta\beta
\end{equation}

For a translation of the pendulum in the $x\textrm{-}y$ plane -- related to the swing motions of the pendulum -- we obtain
\begin{equation}
\label{eq:misalignment}
        \delta s=v\sin(\phi-\alpha)~\delta\alpha-v\beta\cos(\phi-\alpha)~\delta\beta.
\end{equation}
Here $v$ is the magnitude of the vector $\vec{v}$, and $\phi$ is its angle to the $x$ axis. For practical purposes, the second term can be neglected in comparison to the first one, as it contains two factors of small angles. Therefore, we see that if the pendulum is bent ($\delta\alpha\neq0$), the zigzag mode frequency changes in response to angle-conserving translations. The change is maximized to $\delta s=v~\delta\alpha$ for $\phi-\alpha=\pi/2$, i.e. for `transverse' translations of the pendulum, and practically zeroed if $\phi=\alpha$, i.e. for `longitudinal' translations.

\begin{figure}[t]
\centering
    \includegraphics{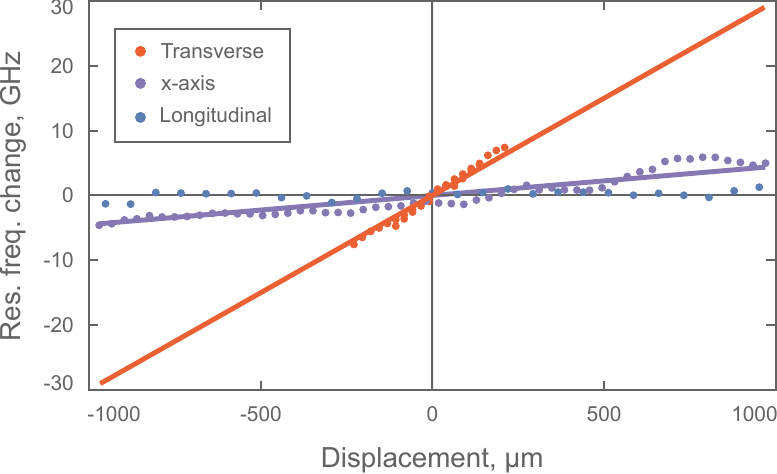}
    \caption{\label{fig:figA31} Zigzag mode frequency change as a function of transverse, x-axis, and longitudinal translations of the pendulum. Solid lines represent the theoretical models based on Eq. \ref{eq:misalignment} assuming $\delta\alpha=0.2$ degrees.}
\end{figure}

Our setup conforms to these expectations as illustrated in Fig. \ref{fig:figA31}, which compares three directions of translations: (1) along the transverse direction (same as Fig. \subref*{fig:fig2_5}), where the fit to the data yields $\delta\alpha=0.2$ degrees; (2) along the $x$ axis ($\phi=0$), in which case the frequency response should be $\sin\alpha=0.15$ times smaller (purple line) than that for the transverse displacements; and (3) along the longitudinal direction, where there is a lack of any pronounced cavity shift. 

For a translation of the pendulum in the $z$ direction, a cavity length change originates from the pitch misalignment $\delta\beta$, yielding $\delta s=v~\delta\beta$. Nevertheless such a change is not very relevant for a real pendulum motion given the constrained motion in the $z$ direction due to the suspension.

We now turn to the roll motion. In presence of finite pitch angles $\beta_{1,2}$, the cavity resonance frequency is expected to develop a residual sensitivity to pure roll rotations (Fig. \ref{fig:fig1}) as well. To model this effect, note that a small roll rotation can be approximated by translations of the two mirror ends in opposite directions along the $z$ axis. This situation can be analyzed by replacing Eq.\ref{eq:projection} by $\delta s=\hat{n}_1\cdot\vec{v}_1+\hat{n}_2\cdot\vec{v}_2$, and using the translation vectors $\vec{v}_{1,2}=\mp \hat{z} \gamma l/2$. Here, $\gamma$ is the small roll angle, and $l$ is the pendulum width as in the main text. With these parameters, the resulting cavity length change can be expressed as $\delta s\approx \gamma l\beta$, showing that the effect is caused by the central angle $\beta$ -- and not $\delta\beta$. To get a sense of the strength of this effect, we can compare it with the measured residual transverse translation sensitivity. Notice that a real transverse swing motion of the pendulum will involve a finite roll motion in addition to pure translations. For a nonzero central pitch angle $\beta$ to cause a leakage signal contribution of the same order as that caused by the translation part of the motion, the undesired pitch angle would need to be of order 0.5 degrees.   

An important takeaway from this analysis is that the relative pitch between the cavity axis and the pendulum should be aligned not only for the sake of minimizing cavity sensitivity to pitch motion, but also to minimize sensitivity to roll motion. One can additionally calculate (based on Eq. \ref{eq:angle_condition}) that the center (i.e., the minimum) of the quadratic cavity response curve to pure pitch rotations remain at the $\beta=0$ condition even in presence of finite yaw and pitch misalignment angles $\delta\alpha$ and $\delta\beta$ (to first order in each of these variables). Thus, to this approximation, there are no conflicting conditions in simultaneously having low sensitivity to any pendulum normal mode that is not the torsional yaw mode -- e.g., transverse or longitudinal swing, roll, pitch, violin, and pendulum bar bending.

Although we argued that a pure roll motion of the pendulum is decoupled from the cavity for $\beta=0$, in reality the physical roll-normal-mode of a pendulum will exhibit finite amounts of transverse translations due to the dynamical coupling between the two degrees of freedom -- because of the constraint introduced by the pendulum suspension \cite{shimoda2018}. We will not reproduce the relevant calculations, but in effect, the roll motion acquires an accompanying transverse translation motion (to first order) by the amount $-\frac{I}{\xi m}\gamma$, where $I$ is the moment of inertia about the center of mass for roll rotations, $\xi$ is the suspension fiber length, and $m$ is the mass of the pendulum body. These small transverse translations form the mechanism for the coupling of the pendulum roll-normal-mode to the cavity when the pendulum is bent (i.e., when $\delta\alpha\neq0$); this is the origin of the leakage roll thermal noise in Fig.\ref{fig:fig4}. For technical completeness, note that the choice of generalized coordinates which leads to the specific minus in the specified translation is when the signs are the same for both translation and roll angles for the swing-normal-mode.

\section{Sensing range}
\label{sec:range}

A limitation to the range of angles that can be sensed comes about due to the change in the geometry of the zigzag mode accompanying a yaw rotation. Here, a sufficiently large change causes the input light to be no longer mode matched to the cavity. We define the limits of the range as the changes in yaw angle $\pm \Delta \theta$ that result in a reduction of the intensity coupling efficiency to $1/e$ times its maximal value.

Calculating the drop in the coupling efficiency amounts to evaluating the overlap between transverse mode profiles of two zigzag modes corresponding to two pendulum orientations with $\Delta \theta$ yaw angle separation. Since paraxial propagation of Gaussian beams preserve the overlap between arbitrary transverse mode profiles, the desired overlap can be calculated at any point along the zigzag mode path -- not necessarily where the light is in-coupled. The center of the cavity -- one of the beam waist locations -- is a particularly convenient location to carry out this calculation, since only the tilt angle $\alpha$ (Fig. \ref{fig:figA1}) of the mode can change at this location in the ideal case. This change is $\Delta \alpha \approx \frac{1}{g} \Delta \theta$ under the assumption $\alpha \ll 1$. We calculate the overlap $|\int_{-\infty}^{\infty} \psi_2^*(x)\, \psi_1(x)\, dx\, |^2$ for the two normalized transverse mode functions $\psi_1(x)=(\frac{2}{\pi \mathrm{w}_0^2})^{1/4} exp(-\frac{x^2}{\mathrm{w}_0^2})$ and $\psi_2(x)=(\frac{2}{\pi \mathrm{w}_0^2})^{1/4} exp(-\frac{x^2}{\mathrm{w}_0^2})exp(i \Delta \alpha \frac{2 \pi}{\lambda} x)$, where $\mathrm{w}_0=\sqrt{\frac{\lambda L}{2 \pi}} (\frac{1+g}{1-g})^{1/4}$ \cite{nagourney2010} is the beam waist size. Carrying out the overlap integral yields
\begin{equation}
    \Delta \theta = g \frac{\lambda}{\pi \mathrm{w}_0}.
\end{equation}
Accounting for rotations of both signs, the full sensing range is given by twice this expression: $\theta_{rng}=2\, \Delta \theta$

\bibliography{td}

\begin{thebibliography}{50}%
\makeatletter
\providecommand \@ifxundefined [1]{%
 \@ifx{#1\undefined}
}%
\providecommand \@ifnum [1]{%
 \ifnum #1\expandafter \@firstoftwo
 \else \expandafter \@secondoftwo
 \fi
}%
\providecommand \@ifx [1]{%
 \ifx #1\expandafter \@firstoftwo
 \else \expandafter \@secondoftwo
 \fi
}%
\providecommand \natexlab [1]{#1}%
\providecommand \enquote  [1]{``#1''}%
\providecommand \bibnamefont  [1]{#1}%
\providecommand \bibfnamefont [1]{#1}%
\providecommand \citenamefont [1]{#1}%
\providecommand \href@noop [0]{\@secondoftwo}%
\providecommand \href [0]{\begingroup \@sanitize@url \@href}%
\providecommand \@href[1]{\@@startlink{#1}\@@href}%
\providecommand \@@href[1]{\endgroup#1\@@endlink}%
\providecommand \@sanitize@url [0]{\catcode `\\12\catcode `\$12\catcode
  `\&12\catcode `\#12\catcode `\^12\catcode `\_12\catcode `\%12\relax}%
\providecommand \@@startlink[1]{}%
\providecommand \@@endlink[0]{}%
\providecommand \url  [0]{\begingroup\@sanitize@url \@url }%
\providecommand \@url [1]{\endgroup\@href {#1}{\urlprefix }}%
\providecommand \urlprefix  [0]{URL }%
\providecommand \Eprint [0]{\href }%
\providecommand \doibase [0]{https://doi.org/}%
\providecommand \selectlanguage [0]{\@gobble}%
\providecommand \bibinfo  [0]{\@secondoftwo}%
\providecommand \bibfield  [0]{\@secondoftwo}%
\providecommand \translation [1]{[#1]}%
\providecommand \BibitemOpen [0]{}%
\providecommand \bibitemStop [0]{}%
\providecommand \bibitemNoStop [0]{.\EOS\space}%
\providecommand \EOS [0]{\spacefactor3000\relax}%
\providecommand \BibitemShut  [1]{\csname bibitem#1\endcsname}%
\let\auto@bib@innerbib\@empty
\bibitem [{\citenamefont {Abbott}\ \emph {et~al.}(2016)\citenamefont {Abbott}
  \emph {et~al.}}]{PhysRevLett.116.061102}%
  \BibitemOpen
  \bibfield  {author} {\bibinfo {author} {\bibfnamefont {B.~P.}\ \bibnamefont
  {Abbott}} \emph {et~al.} (\bibinfo {collaboration} {LIGO Scientific
  Collaboration and Virgo Collaboration}),\ }\bibfield  {title} {\bibinfo
  {title} {Observation of gravitational waves from a binary black hole
  merger},\ }\href {https://doi.org/10.1103/PhysRevLett.116.061102} {\bibfield
  {journal} {\bibinfo  {journal} {Phys. Rev. Lett.}\ }\textbf {\bibinfo
  {volume} {116}},\ \bibinfo {pages} {061102} (\bibinfo {year}
  {2016})}\BibitemShut {NoStop}%
\bibitem [{\citenamefont {Marshall}\ \emph {et~al.}(2003)\citenamefont
  {Marshall}, \citenamefont {Simon}, \citenamefont {Penrose},\ and\
  \citenamefont {Bouwmeester}}]{PhysRevLett.91.130401}%
  \BibitemOpen
  \bibfield  {author} {\bibinfo {author} {\bibfnamefont {W.}~\bibnamefont
  {Marshall}}, \bibinfo {author} {\bibfnamefont {C.}~\bibnamefont {Simon}},
  \bibinfo {author} {\bibfnamefont {R.}~\bibnamefont {Penrose}},\ and\ \bibinfo
  {author} {\bibfnamefont {D.}~\bibnamefont {Bouwmeester}},\ }\bibfield
  {title} {\bibinfo {title} {Towards quantum superpositions of a mirror},\
  }\href {https://doi.org/10.1103/PhysRevLett.91.130401} {\bibfield  {journal}
  {\bibinfo  {journal} {Phys. Rev. Lett.}\ }\textbf {\bibinfo {volume} {91}},\
  \bibinfo {pages} {130401} (\bibinfo {year} {2003})}\BibitemShut {NoStop}%
\bibitem [{\citenamefont {Pikovski}\ \emph {et~al.}(2012)\citenamefont
  {Pikovski}, \citenamefont {Vanner}, \citenamefont {Aspelmeyer}, \citenamefont
  {Kim},\ and\ \citenamefont {Brukner}}]{Pikovski2012}%
  \BibitemOpen
  \bibfield  {author} {\bibinfo {author} {\bibfnamefont {I.}~\bibnamefont
  {Pikovski}}, \bibinfo {author} {\bibfnamefont {M.~R.}\ \bibnamefont
  {Vanner}}, \bibinfo {author} {\bibfnamefont {M.}~\bibnamefont {Aspelmeyer}},
  \bibinfo {author} {\bibfnamefont {M.~S.}\ \bibnamefont {Kim}},\ and\ \bibinfo
  {author} {\bibfnamefont {{\v{C}}.}~\bibnamefont {Brukner}},\ }\bibfield
  {title} {\bibinfo {title} {Probing planck-scale physics with quantum
  optics},\ }\href {https://doi.org/10.1038/nphys2262} {\bibfield  {journal}
  {\bibinfo  {journal} {Nature Physics}\ }\textbf {\bibinfo {volume} {8}},\
  \bibinfo {pages} {393} (\bibinfo {year} {2012})}\BibitemShut {NoStop}%
\bibitem [{\citenamefont {Chan}\ \emph {et~al.}(2011)\citenamefont {Chan},
  \citenamefont {Alegre}, \citenamefont {Safavi-Naeini}, \citenamefont {Hill},
  \citenamefont {Krause}, \citenamefont {Gr{\"o}blacher}, \citenamefont
  {Aspelmeyer},\ and\ \citenamefont {Painter}}]{Chan2011}%
  \BibitemOpen
  \bibfield  {author} {\bibinfo {author} {\bibfnamefont {J.}~\bibnamefont
  {Chan}}, \bibinfo {author} {\bibfnamefont {T.~P.~M.}\ \bibnamefont {Alegre}},
  \bibinfo {author} {\bibfnamefont {A.~H.}\ \bibnamefont {Safavi-Naeini}},
  \bibinfo {author} {\bibfnamefont {J.~T.}\ \bibnamefont {Hill}}, \bibinfo
  {author} {\bibfnamefont {A.}~\bibnamefont {Krause}}, \bibinfo {author}
  {\bibfnamefont {S.}~\bibnamefont {Gr{\"o}blacher}}, \bibinfo {author}
  {\bibfnamefont {M.}~\bibnamefont {Aspelmeyer}},\ and\ \bibinfo {author}
  {\bibfnamefont {O.}~\bibnamefont {Painter}},\ }\bibfield  {title} {\bibinfo
  {title} {Laser cooling of a nanomechanical oscillator into its quantum ground
  state},\ }\href {https://doi.org/10.1038/nature10461} {\bibfield  {journal}
  {\bibinfo  {journal} {Nature}\ }\textbf {\bibinfo {volume} {478}},\ \bibinfo
  {pages} {89} (\bibinfo {year} {2011})}\BibitemShut {NoStop}%
\bibitem [{\citenamefont {Vinante}\ \emph {et~al.}(2020)\citenamefont
  {Vinante}, \citenamefont {Carlesso}, \citenamefont {Bassi}, \citenamefont
  {Chiasera}, \citenamefont {Varas}, \citenamefont {Falferi}, \citenamefont
  {Margesin}, \citenamefont {Mezzena},\ and\ \citenamefont
  {Ulbricht}}]{Vinante2020}%
  \BibitemOpen
  \bibfield  {author} {\bibinfo {author} {\bibfnamefont {A.}~\bibnamefont
  {Vinante}}, \bibinfo {author} {\bibfnamefont {M.}~\bibnamefont {Carlesso}},
  \bibinfo {author} {\bibfnamefont {A.}~\bibnamefont {Bassi}}, \bibinfo
  {author} {\bibfnamefont {A.}~\bibnamefont {Chiasera}}, \bibinfo {author}
  {\bibfnamefont {S.}~\bibnamefont {Varas}}, \bibinfo {author} {\bibfnamefont
  {P.}~\bibnamefont {Falferi}}, \bibinfo {author} {\bibfnamefont
  {B.}~\bibnamefont {Margesin}}, \bibinfo {author} {\bibfnamefont
  {R.}~\bibnamefont {Mezzena}},\ and\ \bibinfo {author} {\bibfnamefont
  {H.}~\bibnamefont {Ulbricht}},\ }\bibfield  {title} {\bibinfo {title}
  {Narrowing the parameter space of collapse models with ultracold layered
  force sensors},\ }\href@noop {} {\bibfield  {journal} {\bibinfo  {journal}
  {Physical Review Letters}\ }\textbf {\bibinfo {volume} {125}},\ \bibinfo
  {pages} {100404} (\bibinfo {year} {2020})}\BibitemShut {NoStop}%
\bibitem [{\citenamefont {Kaltenbaek}\ \emph {et~al.}(2023)\citenamefont
  {Kaltenbaek}, \citenamefont {Arndt}, \citenamefont {Aspelmeyer},
  \citenamefont {Barker}, \citenamefont {Bassi}, \citenamefont {Bateman},
  \citenamefont {Belenchia}, \citenamefont {Berg{\'e}}, \citenamefont
  {Braxmaier}, \citenamefont {Bose} \emph {et~al.}}]{Kaltenbaek2023}%
  \BibitemOpen
  \bibfield  {author} {\bibinfo {author} {\bibfnamefont {R.}~\bibnamefont
  {Kaltenbaek}}, \bibinfo {author} {\bibfnamefont {M.}~\bibnamefont {Arndt}},
  \bibinfo {author} {\bibfnamefont {M.}~\bibnamefont {Aspelmeyer}}, \bibinfo
  {author} {\bibfnamefont {P.~F.}\ \bibnamefont {Barker}}, \bibinfo {author}
  {\bibfnamefont {A.}~\bibnamefont {Bassi}}, \bibinfo {author} {\bibfnamefont
  {J.}~\bibnamefont {Bateman}}, \bibinfo {author} {\bibfnamefont
  {A.}~\bibnamefont {Belenchia}}, \bibinfo {author} {\bibfnamefont
  {J.}~\bibnamefont {Berg{\'e}}}, \bibinfo {author} {\bibfnamefont
  {C.}~\bibnamefont {Braxmaier}}, \bibinfo {author} {\bibfnamefont
  {S.}~\bibnamefont {Bose}}, \emph {et~al.},\ }\bibfield  {title} {\bibinfo
  {title} {Research campaign: Macroscopic quantum resonators (maqro)},\
  }\href@noop {} {\bibfield  {journal} {\bibinfo  {journal} {Quantum Science
  and Technology}\ }\textbf {\bibinfo {volume} {8}},\ \bibinfo {pages} {014006}
  (\bibinfo {year} {2023})}\BibitemShut {NoStop}%
\bibitem [{\citenamefont {Arkani–Hamed}\ \emph {et~al.}(1998)\citenamefont
  {Arkani–Hamed}, \citenamefont {Dimopoulos},\ and\ \citenamefont
  {Dvali}}]{ARKANIHAMED1998263}%
  \BibitemOpen
  \bibfield  {author} {\bibinfo {author} {\bibfnamefont {N.}~\bibnamefont
  {Arkani–Hamed}}, \bibinfo {author} {\bibfnamefont {S.}~\bibnamefont
  {Dimopoulos}},\ and\ \bibinfo {author} {\bibfnamefont {G.}~\bibnamefont
  {Dvali}},\ }\bibfield  {title} {\bibinfo {title} {The hierarchy problem and
  new dimensions at a millimeter},\ }\href
  {https://doi.org/https://doi.org/10.1016/S0370-2693(98)00466-3} {\bibfield
  {journal} {\bibinfo  {journal} {Physics Letters B}\ }\textbf {\bibinfo
  {volume} {429}},\ \bibinfo {pages} {263} (\bibinfo {year}
  {1998})}\BibitemShut {NoStop}%
\bibitem [{\citenamefont {Lee}\ \emph {et~al.}(2020)\citenamefont {Lee},
  \citenamefont {Adelberger}, \citenamefont {Cook}, \citenamefont {Fleischer},\
  and\ \citenamefont {Heckel}}]{PhysRevLett.124.101101}%
  \BibitemOpen
  \bibfield  {author} {\bibinfo {author} {\bibfnamefont {J.~G.}\ \bibnamefont
  {Lee}}, \bibinfo {author} {\bibfnamefont {E.~G.}\ \bibnamefont {Adelberger}},
  \bibinfo {author} {\bibfnamefont {T.~S.}\ \bibnamefont {Cook}}, \bibinfo
  {author} {\bibfnamefont {S.~M.}\ \bibnamefont {Fleischer}},\ and\ \bibinfo
  {author} {\bibfnamefont {B.~R.}\ \bibnamefont {Heckel}},\ }\bibfield  {title}
  {\bibinfo {title} {New test of the gravitational $1/{r}^{2}$ law at
  separations down to $52\text{ }\text{ }\ensuremath{\mu}\mathrm{m}$},\ }\href
  {https://doi.org/10.1103/PhysRevLett.124.101101} {\bibfield  {journal}
  {\bibinfo  {journal} {Phys. Rev. Lett.}\ }\textbf {\bibinfo {volume} {124}},\
  \bibinfo {pages} {101101} (\bibinfo {year} {2020})}\BibitemShut {NoStop}%
\bibitem [{\citenamefont {Westphal}\ \emph {et~al.}(2021)\citenamefont
  {Westphal}, \citenamefont {Hepach}, \citenamefont {Pfaff},\ and\
  \citenamefont {Aspelmeyer}}]{Westphal2021}%
  \BibitemOpen
  \bibfield  {author} {\bibinfo {author} {\bibfnamefont {T.}~\bibnamefont
  {Westphal}}, \bibinfo {author} {\bibfnamefont {H.}~\bibnamefont {Hepach}},
  \bibinfo {author} {\bibfnamefont {J.}~\bibnamefont {Pfaff}},\ and\ \bibinfo
  {author} {\bibfnamefont {M.}~\bibnamefont {Aspelmeyer}},\ }\bibfield  {title}
  {\bibinfo {title} {Measurement of gravitational coupling between
  millimetre-sized masses},\ }\href
  {https://doi.org/10.1038/s41586-021-03250-7} {\bibfield  {journal} {\bibinfo
  {journal} {Nature}\ }\textbf {\bibinfo {volume} {591}},\ \bibinfo {pages}
  {225} (\bibinfo {year} {2021})}\BibitemShut {NoStop}%
\bibitem [{\citenamefont {Spengler}\ \emph {et~al.}(2022)\citenamefont
  {Spengler}, \citenamefont {R{\"a}tzel},\ and\ \citenamefont
  {Braun}}]{Spengler2022}%
  \BibitemOpen
  \bibfield  {author} {\bibinfo {author} {\bibfnamefont {F.}~\bibnamefont
  {Spengler}}, \bibinfo {author} {\bibfnamefont {D.}~\bibnamefont
  {R{\"a}tzel}},\ and\ \bibinfo {author} {\bibfnamefont {D.}~\bibnamefont
  {Braun}},\ }\bibfield  {title} {\bibinfo {title} {Perspectives of measuring
  gravitational effects of laser light and particle beams},\ }\href@noop {}
  {\bibfield  {journal} {\bibinfo  {journal} {New Journal of Physics}\ }\textbf
  {\bibinfo {volume} {24}},\ \bibinfo {pages} {053021} (\bibinfo {year}
  {2022})}\BibitemShut {NoStop}%
\bibitem [{\citenamefont {Gillies}(1997)}]{George_T_Gillies_1997}%
  \BibitemOpen
  \bibfield  {author} {\bibinfo {author} {\bibfnamefont {G.~T.}\ \bibnamefont
  {Gillies}},\ }\bibfield  {title} {\bibinfo {title} {The newtonian
  gravitational constant: recent measurements and related studies},\ }\href
  {https://doi.org/10.1088/0034-4885/60/2/001} {\bibfield  {journal} {\bibinfo
  {journal} {Reports on Progress in Physics}\ }\textbf {\bibinfo {volume}
  {60}},\ \bibinfo {pages} {151} (\bibinfo {year} {1997})}\BibitemShut
  {NoStop}%
\bibitem [{\citenamefont {Komori}\ \emph {et~al.}(2020)\citenamefont {Komori},
  \citenamefont {Enomoto}, \citenamefont {Ooi}, \citenamefont {Miyazaki},
  \citenamefont {Matsumoto}, \citenamefont {Sudhir}, \citenamefont
  {Michimura},\ and\ \citenamefont {Ando}}]{PhysRevA.101.011802}%
  \BibitemOpen
  \bibfield  {author} {\bibinfo {author} {\bibfnamefont {K.}~\bibnamefont
  {Komori}}, \bibinfo {author} {\bibfnamefont {Y.}~\bibnamefont {Enomoto}},
  \bibinfo {author} {\bibfnamefont {C.~P.}\ \bibnamefont {Ooi}}, \bibinfo
  {author} {\bibfnamefont {Y.}~\bibnamefont {Miyazaki}}, \bibinfo {author}
  {\bibfnamefont {N.}~\bibnamefont {Matsumoto}}, \bibinfo {author}
  {\bibfnamefont {V.}~\bibnamefont {Sudhir}}, \bibinfo {author} {\bibfnamefont
  {Y.}~\bibnamefont {Michimura}},\ and\ \bibinfo {author} {\bibfnamefont
  {M.}~\bibnamefont {Ando}},\ }\bibfield  {title} {\bibinfo {title}
  {Attonewton-meter torque sensing with a macroscopic optomechanical torsion
  pendulum},\ }\href {https://doi.org/10.1103/PhysRevA.101.011802} {\bibfield
  {journal} {\bibinfo  {journal} {Phys. Rev. A}\ }\textbf {\bibinfo {volume}
  {101}},\ \bibinfo {pages} {011802} (\bibinfo {year} {2020})}\BibitemShut
  {NoStop}%
\bibitem [{\citenamefont {Bushev}\ \emph {et~al.}(2019)\citenamefont {Bushev},
  \citenamefont {Bourhill}, \citenamefont {Goryachev}, \citenamefont
  {Kukharchyk}, \citenamefont {Ivanov}, \citenamefont {Galliou}, \citenamefont
  {Tobar},\ and\ \citenamefont {Danilishin}}]{PhysRevD.100.066020}%
  \BibitemOpen
  \bibfield  {author} {\bibinfo {author} {\bibfnamefont {P.~A.}\ \bibnamefont
  {Bushev}}, \bibinfo {author} {\bibfnamefont {J.}~\bibnamefont {Bourhill}},
  \bibinfo {author} {\bibfnamefont {M.}~\bibnamefont {Goryachev}}, \bibinfo
  {author} {\bibfnamefont {N.}~\bibnamefont {Kukharchyk}}, \bibinfo {author}
  {\bibfnamefont {E.}~\bibnamefont {Ivanov}}, \bibinfo {author} {\bibfnamefont
  {S.}~\bibnamefont {Galliou}}, \bibinfo {author} {\bibfnamefont {M.~E.}\
  \bibnamefont {Tobar}},\ and\ \bibinfo {author} {\bibfnamefont
  {S.}~\bibnamefont {Danilishin}},\ }\bibfield  {title} {\bibinfo {title}
  {Testing the generalized uncertainty principle with macroscopic mechanical
  oscillators and pendulums},\ }\href
  {https://doi.org/10.1103/PhysRevD.100.066020} {\bibfield  {journal} {\bibinfo
   {journal} {Phys. Rev. D}\ }\textbf {\bibinfo {volume} {100}},\ \bibinfo
  {pages} {066020} (\bibinfo {year} {2019})}\BibitemShut {NoStop}%
\bibitem [{\citenamefont {Altin}\ \emph {et~al.}(2017)\citenamefont {Altin},
  \citenamefont {Nguyen}, \citenamefont {Slagmolen}, \citenamefont {Ward},
  \citenamefont {Shaddock},\ and\ \citenamefont {McClelland}}]{Altin2017}%
  \BibitemOpen
  \bibfield  {author} {\bibinfo {author} {\bibfnamefont {P.~A.}\ \bibnamefont
  {Altin}}, \bibinfo {author} {\bibfnamefont {T.~T.-H.}\ \bibnamefont
  {Nguyen}}, \bibinfo {author} {\bibfnamefont {B.~J.~J.}\ \bibnamefont
  {Slagmolen}}, \bibinfo {author} {\bibfnamefont {R.~L.}\ \bibnamefont {Ward}},
  \bibinfo {author} {\bibfnamefont {D.~A.}\ \bibnamefont {Shaddock}},\ and\
  \bibinfo {author} {\bibfnamefont {D.~E.}\ \bibnamefont {McClelland}},\
  }\bibfield  {title} {\bibinfo {title} {A robust single-beam optical trap for
  a gram-scale mechanical oscillator},\ }\href
  {https://doi.org/10.1038/s41598-017-15179-x} {\bibfield  {journal} {\bibinfo
  {journal} {Scientific Reports}\ }\textbf {\bibinfo {volume} {7}},\ \bibinfo
  {pages} {14546} (\bibinfo {year} {2017})}\BibitemShut {NoStop}%
\bibitem [{\citenamefont {Matsumoto}\ \emph {et~al.}(2019)\citenamefont
  {Matsumoto}, \citenamefont {Cata{\~n}o-Lopez}, \citenamefont {Sugawara},
  \citenamefont {Suzuki}, \citenamefont {Abe}, \citenamefont {Komori},
  \citenamefont {Michimura}, \citenamefont {Aso},\ and\ \citenamefont
  {Edamatsu}}]{Matsumoto2019}%
  \BibitemOpen
  \bibfield  {author} {\bibinfo {author} {\bibfnamefont {N.}~\bibnamefont
  {Matsumoto}}, \bibinfo {author} {\bibfnamefont {S.~B.}\ \bibnamefont
  {Cata{\~n}o-Lopez}}, \bibinfo {author} {\bibfnamefont {M.}~\bibnamefont
  {Sugawara}}, \bibinfo {author} {\bibfnamefont {S.}~\bibnamefont {Suzuki}},
  \bibinfo {author} {\bibfnamefont {N.}~\bibnamefont {Abe}}, \bibinfo {author}
  {\bibfnamefont {K.}~\bibnamefont {Komori}}, \bibinfo {author} {\bibfnamefont
  {Y.}~\bibnamefont {Michimura}}, \bibinfo {author} {\bibfnamefont
  {Y.}~\bibnamefont {Aso}},\ and\ \bibinfo {author} {\bibfnamefont
  {K.}~\bibnamefont {Edamatsu}},\ }\bibfield  {title} {\bibinfo {title}
  {Demonstration of displacement sensing of a mg-scale pendulum for mm-and
  mg-scale gravity measurements},\ }\href@noop {} {\bibfield  {journal}
  {\bibinfo  {journal} {Physical review letters}\ }\textbf {\bibinfo {volume}
  {122}},\ \bibinfo {pages} {071101} (\bibinfo {year} {2019})}\BibitemShut
  {NoStop}%
\bibitem [{\citenamefont {Cata{\~n}o-Lopez}\ \emph {et~al.}(2020)\citenamefont
  {Cata{\~n}o-Lopez}, \citenamefont {Santiago-Condori}, \citenamefont
  {Edamatsu},\ and\ \citenamefont {Matsumoto}}]{Catano2020}%
  \BibitemOpen
  \bibfield  {author} {\bibinfo {author} {\bibfnamefont {S.~B.}\ \bibnamefont
  {Cata{\~n}o-Lopez}}, \bibinfo {author} {\bibfnamefont {J.~G.}\ \bibnamefont
  {Santiago-Condori}}, \bibinfo {author} {\bibfnamefont {K.}~\bibnamefont
  {Edamatsu}},\ and\ \bibinfo {author} {\bibfnamefont {N.}~\bibnamefont
  {Matsumoto}},\ }\bibfield  {title} {\bibinfo {title} {High-q milligram-scale
  monolithic pendulum for quantum-limited gravity measurements},\ }\href@noop
  {} {\bibfield  {journal} {\bibinfo  {journal} {Physical review letters}\
  }\textbf {\bibinfo {volume} {124}},\ \bibinfo {pages} {221102} (\bibinfo
  {year} {2020})}\BibitemShut {NoStop}%
\bibitem [{\citenamefont {Michimura}\ and\ \citenamefont
  {Komori}(2020)}]{Michimura2020}%
  \BibitemOpen
  \bibfield  {author} {\bibinfo {author} {\bibfnamefont {Y.}~\bibnamefont
  {Michimura}}\ and\ \bibinfo {author} {\bibfnamefont {K.}~\bibnamefont
  {Komori}},\ }\bibfield  {title} {\bibinfo {title} {Quantum sensing with
  milligram scale optomechanical systems},\ }\href
  {https://doi.org/10.1140/epjd/e2020-10185-5} {\bibfield  {journal} {\bibinfo
  {journal} {The European Physical Journal D}\ }\textbf {\bibinfo {volume}
  {74}},\ \bibinfo {pages} {126} (\bibinfo {year} {2020})}\BibitemShut
  {NoStop}%
\bibitem [{\citenamefont {Bose}\ \emph {et~al.}(2017)\citenamefont {Bose},
  \citenamefont {Mazumdar}, \citenamefont {Morley}, \citenamefont {Ulbricht},
  \citenamefont {Toro{\v{s}}}, \citenamefont {Paternostro}, \citenamefont
  {Geraci}, \citenamefont {Barker}, \citenamefont {Kim},\ and\ \citenamefont
  {Milburn}}]{Bose2017}%
  \BibitemOpen
  \bibfield  {author} {\bibinfo {author} {\bibfnamefont {S.}~\bibnamefont
  {Bose}}, \bibinfo {author} {\bibfnamefont {A.}~\bibnamefont {Mazumdar}},
  \bibinfo {author} {\bibfnamefont {G.~W.}\ \bibnamefont {Morley}}, \bibinfo
  {author} {\bibfnamefont {H.}~\bibnamefont {Ulbricht}}, \bibinfo {author}
  {\bibfnamefont {M.}~\bibnamefont {Toro{\v{s}}}}, \bibinfo {author}
  {\bibfnamefont {M.}~\bibnamefont {Paternostro}}, \bibinfo {author}
  {\bibfnamefont {A.~A.}\ \bibnamefont {Geraci}}, \bibinfo {author}
  {\bibfnamefont {P.~F.}\ \bibnamefont {Barker}}, \bibinfo {author}
  {\bibfnamefont {M.}~\bibnamefont {Kim}},\ and\ \bibinfo {author}
  {\bibfnamefont {G.}~\bibnamefont {Milburn}},\ }\bibfield  {title} {\bibinfo
  {title} {Spin entanglement witness for quantum gravity},\ }\href@noop {}
  {\bibfield  {journal} {\bibinfo  {journal} {Physical review letters}\
  }\textbf {\bibinfo {volume} {119}},\ \bibinfo {pages} {240401} (\bibinfo
  {year} {2017})}\BibitemShut {NoStop}%
\bibitem [{\citenamefont {Marletto}\ and\ \citenamefont
  {Vedral}(2017)}]{Marletto2017}%
  \BibitemOpen
  \bibfield  {author} {\bibinfo {author} {\bibfnamefont {C.}~\bibnamefont
  {Marletto}}\ and\ \bibinfo {author} {\bibfnamefont {V.}~\bibnamefont
  {Vedral}},\ }\bibfield  {title} {\bibinfo {title} {Gravitationally induced
  entanglement between two massive particles is sufficient evidence of quantum
  effects in gravity},\ }\href@noop {} {\bibfield  {journal} {\bibinfo
  {journal} {Physical review letters}\ }\textbf {\bibinfo {volume} {119}},\
  \bibinfo {pages} {240402} (\bibinfo {year} {2017})}\BibitemShut {NoStop}%
\bibitem [{\citenamefont {Krisnanda}\ \emph {et~al.}(2020)\citenamefont
  {Krisnanda}, \citenamefont {Tham}, \citenamefont {Paternostro},\ and\
  \citenamefont {Paterek}}]{Krisnanda2020}%
  \BibitemOpen
  \bibfield  {author} {\bibinfo {author} {\bibfnamefont {T.}~\bibnamefont
  {Krisnanda}}, \bibinfo {author} {\bibfnamefont {G.~Y.}\ \bibnamefont {Tham}},
  \bibinfo {author} {\bibfnamefont {M.}~\bibnamefont {Paternostro}},\ and\
  \bibinfo {author} {\bibfnamefont {T.}~\bibnamefont {Paterek}},\ }\bibfield
  {title} {\bibinfo {title} {Observable quantum entanglement due to gravity},\
  }\href {https://doi.org/10.1038/s41534-020-0243-y} {\bibfield  {journal}
  {\bibinfo  {journal} {npj Quantum Information}\ }\textbf {\bibinfo {volume}
  {6}},\ \bibinfo {pages} {12} (\bibinfo {year} {2020})}\BibitemShut {NoStop}%
\bibitem [{\citenamefont {Carney}\ \emph {et~al.}(2021)\citenamefont {Carney}
  \emph {et~al.}}]{Carney_2021}%
  \BibitemOpen
  \bibfield  {author} {\bibinfo {author} {\bibfnamefont {D.}~\bibnamefont
  {Carney}} \emph {et~al.},\ }\bibfield  {title} {\bibinfo {title} {Mechanical
  quantum sensing in the search for dark matter},\ }\href
  {https://doi.org/10.1088/2058-9565/abcfcd} {\bibfield  {journal} {\bibinfo
  {journal} {Quantum Science and Technology}\ }\textbf {\bibinfo {volume}
  {6}},\ \bibinfo {pages} {024002} (\bibinfo {year} {2021})}\BibitemShut
  {NoStop}%
\bibitem [{\citenamefont {Whittle}\ \emph {et~al.}(2021)\citenamefont {Whittle}
  \emph {et~al.}}]{doi:10.1126/science.abh2634}%
  \BibitemOpen
  \bibfield  {author} {\bibinfo {author} {\bibfnamefont {C.}~\bibnamefont
  {Whittle}} \emph {et~al.},\ }\bibfield  {title} {\bibinfo {title}
  {Approaching the motional ground state of a 10-kg object},\ }\href
  {https://doi.org/10.1126/science.abh2634} {\bibfield  {journal} {\bibinfo
  {journal} {Science}\ }\textbf {\bibinfo {volume} {372}},\ \bibinfo {pages}
  {1333} (\bibinfo {year} {2021})},\ \Eprint
  {https://arxiv.org/abs/https://www.science.org/doi/pdf/10.1126/science.abh2634}
  {https://www.science.org/doi/pdf/10.1126/science.abh2634} \BibitemShut
  {NoStop}%
\bibitem [{\citenamefont {von E{\"o}tv{\"o}s}(1890)}]{von1890mathematische}%
  \BibitemOpen
  \bibfield  {author} {\bibinfo {author} {\bibfnamefont {R.}~\bibnamefont {von
  E{\"o}tv{\"o}s}},\ }\bibfield  {title} {\bibinfo {title} {Mathematische und
  naturwissenschaftliche berichte aus ungarn},\ }\href@noop {} {\bibfield
  {journal} {\bibinfo  {journal} {Bd}\ }\textbf {\bibinfo {volume} {8}},\
  \bibinfo {pages} {65} (\bibinfo {year} {1890})}\BibitemShut {NoStop}%
\bibitem [{\citenamefont {Alexander}\ \emph {et~al.}(1989)\citenamefont
  {Alexander}, \citenamefont {Hellemans}, \citenamefont {Marti}, \citenamefont
  {Schneir}, \citenamefont {Elings}, \citenamefont {Hansma}, \citenamefont
  {Longmire},\ and\ \citenamefont {Gurley}}]{Alexander1989}%
  \BibitemOpen
  \bibfield  {author} {\bibinfo {author} {\bibfnamefont {S.}~\bibnamefont
  {Alexander}}, \bibinfo {author} {\bibfnamefont {L.}~\bibnamefont
  {Hellemans}}, \bibinfo {author} {\bibfnamefont {O.}~\bibnamefont {Marti}},
  \bibinfo {author} {\bibfnamefont {J.}~\bibnamefont {Schneir}}, \bibinfo
  {author} {\bibfnamefont {V.}~\bibnamefont {Elings}}, \bibinfo {author}
  {\bibfnamefont {P.~K.}\ \bibnamefont {Hansma}}, \bibinfo {author}
  {\bibfnamefont {M.}~\bibnamefont {Longmire}},\ and\ \bibinfo {author}
  {\bibfnamefont {J.}~\bibnamefont {Gurley}},\ }\bibfield  {title} {\bibinfo
  {title} {An atomic-resolution atomic-force microscope implemented using an
  optical lever},\ }\href@noop {} {\bibfield  {journal} {\bibinfo  {journal}
  {Journal of applied physics}\ }\textbf {\bibinfo {volume} {65}},\ \bibinfo
  {pages} {164} (\bibinfo {year} {1989})}\BibitemShut {NoStop}%
\bibitem [{\citenamefont {Pratt}\ \emph {et~al.}(2023)\citenamefont {Pratt},
  \citenamefont {Agrawal}, \citenamefont {Condos}, \citenamefont {Pluchar},
  \citenamefont {Schlamminger},\ and\ \citenamefont {Wilson}}]{Pratt2023}%
  \BibitemOpen
  \bibfield  {author} {\bibinfo {author} {\bibfnamefont {J.~R.}\ \bibnamefont
  {Pratt}}, \bibinfo {author} {\bibfnamefont {A.~R.}\ \bibnamefont {Agrawal}},
  \bibinfo {author} {\bibfnamefont {C.~A.}\ \bibnamefont {Condos}}, \bibinfo
  {author} {\bibfnamefont {C.~M.}\ \bibnamefont {Pluchar}}, \bibinfo {author}
  {\bibfnamefont {S.}~\bibnamefont {Schlamminger}},\ and\ \bibinfo {author}
  {\bibfnamefont {D.~J.}\ \bibnamefont {Wilson}},\ }\bibfield  {title}
  {\bibinfo {title} {Nanoscale torsional dissipation dilution for quantum
  experiments and precision measurement},\ }\href@noop {} {\bibfield  {journal}
  {\bibinfo  {journal} {Physical Review X}\ }\textbf {\bibinfo {volume} {13}},\
  \bibinfo {pages} {011018} (\bibinfo {year} {2023})}\BibitemShut {NoStop}%
\bibitem [{\citenamefont {Hao}\ and\ \citenamefont
  {Purdy}(2022)}]{https://doi.org/10.48550/arxiv.2212.08197}%
  \BibitemOpen
  \bibfield  {author} {\bibinfo {author} {\bibfnamefont {S.}~\bibnamefont
  {Hao}}\ and\ \bibinfo {author} {\bibfnamefont {T.}~\bibnamefont {Purdy}},\
  }\href {https://doi.org/10.48550/ARXIV.2212.08197} {\bibinfo {title} {Back
  action evasion in optical lever detection}} (\bibinfo {year}
  {2022})\BibitemShut {NoStop}%
\bibitem [{\citenamefont {Park}\ and\ \citenamefont {Cho}(2016)}]{Park:16}%
  \BibitemOpen
  \bibfield  {author} {\bibinfo {author} {\bibfnamefont {J.~G.}\ \bibnamefont
  {Park}}\ and\ \bibinfo {author} {\bibfnamefont {K.}~\bibnamefont {Cho}},\
  }\bibfield  {title} {\bibinfo {title} {High-precision tilt sensor using a
  folded mach--zehnder geometry in-phase and quadrature interferometer},\
  }\href {https://doi.org/10.1364/AO.55.002155} {\bibfield  {journal} {\bibinfo
   {journal} {Appl. Opt.}\ }\textbf {\bibinfo {volume} {55}},\ \bibinfo {pages}
  {2155} (\bibinfo {year} {2016})}\BibitemShut {NoStop}%
\bibitem [{\citenamefont {Ross}\ \emph {et~al.}(2021)\citenamefont {Ross},
  \citenamefont {Venkateswara}, \citenamefont {Hagedorn}, \citenamefont
  {Leupold}, \citenamefont {Forsyth}, \citenamefont {Wegner}, \citenamefont
  {Shaw}, \citenamefont {Lee},\ and\ \citenamefont
  {Gundlach}}]{doi:10.1063/5.0043098}%
  \BibitemOpen
  \bibfield  {author} {\bibinfo {author} {\bibfnamefont {M.~P.}\ \bibnamefont
  {Ross}}, \bibinfo {author} {\bibfnamefont {K.}~\bibnamefont {Venkateswara}},
  \bibinfo {author} {\bibfnamefont {C.~A.}\ \bibnamefont {Hagedorn}}, \bibinfo
  {author} {\bibfnamefont {C.~J.}\ \bibnamefont {Leupold}}, \bibinfo {author}
  {\bibfnamefont {P.~W.~F.}\ \bibnamefont {Forsyth}}, \bibinfo {author}
  {\bibfnamefont {J.~D.}\ \bibnamefont {Wegner}}, \bibinfo {author}
  {\bibfnamefont {E.~A.}\ \bibnamefont {Shaw}}, \bibinfo {author}
  {\bibfnamefont {J.~G.}\ \bibnamefont {Lee}},\ and\ \bibinfo {author}
  {\bibfnamefont {J.~H.}\ \bibnamefont {Gundlach}},\ }\bibfield  {title}
  {\bibinfo {title} {A low-frequency torsion pendulum with interferometric
  readout},\ }\href {https://doi.org/10.1063/5.0043098} {\bibfield  {journal}
  {\bibinfo  {journal} {Review of Scientific Instruments}\ }\textbf {\bibinfo
  {volume} {92}},\ \bibinfo {pages} {054502} (\bibinfo {year} {2021})},\
  \Eprint {https://arxiv.org/abs/https://doi.org/10.1063/5.0043098}
  {https://doi.org/10.1063/5.0043098} \BibitemShut {NoStop}%
\bibitem [{\citenamefont {Smetana}\ \emph {et~al.}(2022)\citenamefont
  {Smetana}, \citenamefont {Walters}, \citenamefont {Bauchinger}, \citenamefont
  {Ubhi}, \citenamefont {Cooper}, \citenamefont {Hoyland}, \citenamefont
  {Abbott}, \citenamefont {Baune}, \citenamefont {Fritchel}, \citenamefont
  {Gerberding} \emph {et~al.}}]{Smetana2022}%
  \BibitemOpen
  \bibfield  {author} {\bibinfo {author} {\bibfnamefont {J.}~\bibnamefont
  {Smetana}}, \bibinfo {author} {\bibfnamefont {R.}~\bibnamefont {Walters}},
  \bibinfo {author} {\bibfnamefont {S.}~\bibnamefont {Bauchinger}}, \bibinfo
  {author} {\bibfnamefont {A.~S.}\ \bibnamefont {Ubhi}}, \bibinfo {author}
  {\bibfnamefont {S.}~\bibnamefont {Cooper}}, \bibinfo {author} {\bibfnamefont
  {D.}~\bibnamefont {Hoyland}}, \bibinfo {author} {\bibfnamefont
  {R.}~\bibnamefont {Abbott}}, \bibinfo {author} {\bibfnamefont
  {C.}~\bibnamefont {Baune}}, \bibinfo {author} {\bibfnamefont
  {P.}~\bibnamefont {Fritchel}}, \bibinfo {author} {\bibfnamefont
  {O.}~\bibnamefont {Gerberding}}, \emph {et~al.},\ }\bibfield  {title}
  {\bibinfo {title} {Compact michelson interferometers with subpicometer
  sensitivity},\ }\href@noop {} {\bibfield  {journal} {\bibinfo  {journal}
  {Physical Review Applied}\ }\textbf {\bibinfo {volume} {18}},\ \bibinfo
  {pages} {034040} (\bibinfo {year} {2022})}\BibitemShut {NoStop}%
\bibitem [{\citenamefont {Mart\'{i}nez-Rinc\'{o}n}\ \emph
  {et~al.}(2017)\citenamefont {Mart\'{i}nez-Rinc\'{o}n}, \citenamefont
  {Mullarkey}, \citenamefont {Viza}, \citenamefont {Liu},\ and\ \citenamefont
  {Howell}}]{Martinez-Rincon:17}%
  \BibitemOpen
  \bibfield  {author} {\bibinfo {author} {\bibfnamefont {J.}~\bibnamefont
  {Mart\'{i}nez-Rinc\'{o}n}}, \bibinfo {author} {\bibfnamefont {C.~A.}\
  \bibnamefont {Mullarkey}}, \bibinfo {author} {\bibfnamefont {G.~I.}\
  \bibnamefont {Viza}}, \bibinfo {author} {\bibfnamefont {W.-T.}\ \bibnamefont
  {Liu}},\ and\ \bibinfo {author} {\bibfnamefont {J.~C.}\ \bibnamefont
  {Howell}},\ }\bibfield  {title} {\bibinfo {title} {Ultrasensitive inverse
  weak-value tilt meter},\ }\href {https://doi.org/10.1364/OL.42.002479}
  {\bibfield  {journal} {\bibinfo  {journal} {Opt. Lett.}\ }\textbf {\bibinfo
  {volume} {42}},\ \bibinfo {pages} {2479} (\bibinfo {year}
  {2017})}\BibitemShut {NoStop}%
\bibitem [{\citenamefont {Hogan}\ \emph {et~al.}(2011)\citenamefont {Hogan},
  \citenamefont {Hammer}, \citenamefont {Chiow}, \citenamefont {Dickerson},
  \citenamefont {Johnson}, \citenamefont {Kovachy}, \citenamefont
  {Sugarbaker},\ and\ \citenamefont {Kasevich}}]{Hogan2011}%
  \BibitemOpen
  \bibfield  {author} {\bibinfo {author} {\bibfnamefont {J.}~\bibnamefont
  {Hogan}}, \bibinfo {author} {\bibfnamefont {J.}~\bibnamefont {Hammer}},
  \bibinfo {author} {\bibfnamefont {S.-W.}\ \bibnamefont {Chiow}}, \bibinfo
  {author} {\bibfnamefont {S.}~\bibnamefont {Dickerson}}, \bibinfo {author}
  {\bibfnamefont {D.}~\bibnamefont {Johnson}}, \bibinfo {author} {\bibfnamefont
  {T.}~\bibnamefont {Kovachy}}, \bibinfo {author} {\bibfnamefont
  {A.}~\bibnamefont {Sugarbaker}},\ and\ \bibinfo {author} {\bibfnamefont
  {M.}~\bibnamefont {Kasevich}},\ }\bibfield  {title} {\bibinfo {title}
  {Precision angle sensor using an optical lever inside a sagnac
  interferometer},\ }\href@noop {} {\bibfield  {journal} {\bibinfo  {journal}
  {Optics letters}\ }\textbf {\bibinfo {volume} {36}},\ \bibinfo {pages} {1698}
  (\bibinfo {year} {2011})}\BibitemShut {NoStop}%
\bibitem [{\citenamefont {Cripe}\ \emph {et~al.}(2019)\citenamefont {Cripe},
  \citenamefont {Aggarwal}, \citenamefont {Lanza}, \citenamefont {Libson},
  \citenamefont {Singh}, \citenamefont {Heu}, \citenamefont {Follman},
  \citenamefont {Cole}, \citenamefont {Mavalvala},\ and\ \citenamefont
  {Corbitt}}]{Cripe2019}%
  \BibitemOpen
  \bibfield  {author} {\bibinfo {author} {\bibfnamefont {J.}~\bibnamefont
  {Cripe}}, \bibinfo {author} {\bibfnamefont {N.}~\bibnamefont {Aggarwal}},
  \bibinfo {author} {\bibfnamefont {R.}~\bibnamefont {Lanza}}, \bibinfo
  {author} {\bibfnamefont {A.}~\bibnamefont {Libson}}, \bibinfo {author}
  {\bibfnamefont {R.}~\bibnamefont {Singh}}, \bibinfo {author} {\bibfnamefont
  {P.}~\bibnamefont {Heu}}, \bibinfo {author} {\bibfnamefont {D.}~\bibnamefont
  {Follman}}, \bibinfo {author} {\bibfnamefont {G.~D.}\ \bibnamefont {Cole}},
  \bibinfo {author} {\bibfnamefont {N.}~\bibnamefont {Mavalvala}},\ and\
  \bibinfo {author} {\bibfnamefont {T.}~\bibnamefont {Corbitt}},\ }\bibfield
  {title} {\bibinfo {title} {Measurement of quantum back action in the audio
  band at room temperature},\ }\href
  {https://doi.org/10.1038/s41586-019-1051-4} {\bibfield  {journal} {\bibinfo
  {journal} {Nature}\ }\textbf {\bibinfo {volume} {568}},\ \bibinfo {pages}
  {364} (\bibinfo {year} {2019})}\BibitemShut {NoStop}%
\bibitem [{\citenamefont {Mueller}\ \emph {et~al.}(2008)\citenamefont
  {Mueller}, \citenamefont {Heugel},\ and\ \citenamefont {Wang}}]{mueller2008}%
  \BibitemOpen
  \bibfield  {author} {\bibinfo {author} {\bibfnamefont {F.}~\bibnamefont
  {Mueller}}, \bibinfo {author} {\bibfnamefont {S.}~\bibnamefont {Heugel}},\
  and\ \bibinfo {author} {\bibfnamefont {L.}~\bibnamefont {Wang}},\ }\bibfield
  {title} {\bibinfo {title} {Observation of optomechanical multistability in a
  high-q torsion balance oscillator},\ }\href@noop {} {\bibfield  {journal}
  {\bibinfo  {journal} {Physical Review A}\ }\textbf {\bibinfo {volume} {77}},\
  \bibinfo {pages} {031802} (\bibinfo {year} {2008})}\BibitemShut {NoStop}%
\bibitem [{\citenamefont {McManus}\ \emph {et~al.}(2017)\citenamefont
  {McManus}, \citenamefont {Forsyth}, \citenamefont {Yap}, \citenamefont
  {Ward}, \citenamefont {Shaddock}, \citenamefont {McClelland},\ and\
  \citenamefont {Slagmolen}}]{mcmanus2017}%
  \BibitemOpen
  \bibfield  {author} {\bibinfo {author} {\bibfnamefont {D.}~\bibnamefont
  {McManus}}, \bibinfo {author} {\bibfnamefont {P.}~\bibnamefont {Forsyth}},
  \bibinfo {author} {\bibfnamefont {M.~J.}\ \bibnamefont {Yap}}, \bibinfo
  {author} {\bibfnamefont {R.}~\bibnamefont {Ward}}, \bibinfo {author}
  {\bibfnamefont {D.}~\bibnamefont {Shaddock}}, \bibinfo {author}
  {\bibfnamefont {D.}~\bibnamefont {McClelland}},\ and\ \bibinfo {author}
  {\bibfnamefont {B.}~\bibnamefont {Slagmolen}},\ }\bibfield  {title} {\bibinfo
  {title} {Mechanical characterisation of the torpedo: a low frequency
  gravitational force sensor},\ }\href@noop {} {\bibfield  {journal} {\bibinfo
  {journal} {Classical and Quantum Gravity}\ }\textbf {\bibinfo {volume}
  {34}},\ \bibinfo {pages} {135002} (\bibinfo {year} {2017})}\BibitemShut
  {NoStop}%
\bibitem [{\citenamefont {Shimoda}\ \emph {et~al.}(2022)\citenamefont
  {Shimoda}, \citenamefont {Miyazaki}, \citenamefont {Enomoto}, \citenamefont
  {Nagano},\ and\ \citenamefont {Ando}}]{Shimoda:22}%
  \BibitemOpen
  \bibfield  {author} {\bibinfo {author} {\bibfnamefont {T.}~\bibnamefont
  {Shimoda}}, \bibinfo {author} {\bibfnamefont {Y.}~\bibnamefont {Miyazaki}},
  \bibinfo {author} {\bibfnamefont {Y.}~\bibnamefont {Enomoto}}, \bibinfo
  {author} {\bibfnamefont {K.}~\bibnamefont {Nagano}},\ and\ \bibinfo {author}
  {\bibfnamefont {M.}~\bibnamefont {Ando}},\ }\bibfield  {title} {\bibinfo
  {title} {Coherent angular signal amplification using an optical cavity},\
  }\href {https://doi.org/10.1364/AO.455485} {\bibfield  {journal} {\bibinfo
  {journal} {Appl. Opt.}\ }\textbf {\bibinfo {volume} {61}},\ \bibinfo {pages}
  {3901} (\bibinfo {year} {2022})}\BibitemShut {NoStop}%
\bibitem [{\citenamefont {Matsumoto}\ \emph {et~al.}(2014)\citenamefont
  {Matsumoto}, \citenamefont {Michimura}, \citenamefont {Aso},\ and\
  \citenamefont {Tsubono}}]{Matsumoto2014}%
  \BibitemOpen
  \bibfield  {author} {\bibinfo {author} {\bibfnamefont {N.}~\bibnamefont
  {Matsumoto}}, \bibinfo {author} {\bibfnamefont {Y.}~\bibnamefont
  {Michimura}}, \bibinfo {author} {\bibfnamefont {Y.}~\bibnamefont {Aso}},\
  and\ \bibinfo {author} {\bibfnamefont {K.}~\bibnamefont {Tsubono}},\
  }\bibfield  {title} {\bibinfo {title} {Optically trapped mirror for reaching
  the standard quantum limit},\ }\href@noop {} {\bibfield  {journal} {\bibinfo
  {journal} {Optics Express}\ }\textbf {\bibinfo {volume} {22}},\ \bibinfo
  {pages} {12915} (\bibinfo {year} {2014})}\BibitemShut {NoStop}%
\bibitem [{\citenamefont {Kuhn}\ \emph {et~al.}(2015)\citenamefont {Kuhn},
  \citenamefont {Asenbaum}, \citenamefont {Kosloff}, \citenamefont {Sclafani},
  \citenamefont {Stickler}, \citenamefont {Nimmrichter}, \citenamefont
  {Hornberger}, \citenamefont {Cheshnovsky}, \citenamefont {Patolsky},\ and\
  \citenamefont {Arndt}}]{Kuhn2015}%
  \BibitemOpen
  \bibfield  {author} {\bibinfo {author} {\bibfnamefont {S.}~\bibnamefont
  {Kuhn}}, \bibinfo {author} {\bibfnamefont {P.}~\bibnamefont {Asenbaum}},
  \bibinfo {author} {\bibfnamefont {A.}~\bibnamefont {Kosloff}}, \bibinfo
  {author} {\bibfnamefont {M.}~\bibnamefont {Sclafani}}, \bibinfo {author}
  {\bibfnamefont {B.~A.}\ \bibnamefont {Stickler}}, \bibinfo {author}
  {\bibfnamefont {S.}~\bibnamefont {Nimmrichter}}, \bibinfo {author}
  {\bibfnamefont {K.}~\bibnamefont {Hornberger}}, \bibinfo {author}
  {\bibfnamefont {O.}~\bibnamefont {Cheshnovsky}}, \bibinfo {author}
  {\bibfnamefont {F.}~\bibnamefont {Patolsky}},\ and\ \bibinfo {author}
  {\bibfnamefont {M.}~\bibnamefont {Arndt}},\ }\bibfield  {title} {\bibinfo
  {title} {Cavity-assisted manipulation of freely rotating silicon nanorods in
  high vacuum},\ }\href@noop {} {\bibfield  {journal} {\bibinfo  {journal}
  {Nano letters}\ }\textbf {\bibinfo {volume} {15}},\ \bibinfo {pages} {5604}
  (\bibinfo {year} {2015})}\BibitemShut {NoStop}%
\bibitem [{\citenamefont {Miao}\ \emph {et~al.}(2020)\citenamefont {Miao},
  \citenamefont {Martynov}, \citenamefont {Yang},\ and\ \citenamefont
  {Datta}}]{PhysRevA.101.063804}%
  \BibitemOpen
  \bibfield  {author} {\bibinfo {author} {\bibfnamefont {H.}~\bibnamefont
  {Miao}}, \bibinfo {author} {\bibfnamefont {D.}~\bibnamefont {Martynov}},
  \bibinfo {author} {\bibfnamefont {H.}~\bibnamefont {Yang}},\ and\ \bibinfo
  {author} {\bibfnamefont {A.}~\bibnamefont {Datta}},\ }\bibfield  {title}
  {\bibinfo {title} {Quantum correlations of light mediated by gravity},\
  }\href {https://doi.org/10.1103/PhysRevA.101.063804} {\bibfield  {journal}
  {\bibinfo  {journal} {Phys. Rev. A}\ }\textbf {\bibinfo {volume} {101}},\
  \bibinfo {pages} {063804} (\bibinfo {year} {2020})}\BibitemShut {NoStop}%
\bibitem [{\citenamefont {Vitali}\ \emph {et~al.}(2007)\citenamefont {Vitali},
  \citenamefont {Gigan}, \citenamefont {Ferreira}, \citenamefont {B\"ohm},
  \citenamefont {Tombesi}, \citenamefont {Guerreiro}, \citenamefont {Vedral},
  \citenamefont {Zeilinger},\ and\ \citenamefont
  {Aspelmeyer}}]{PhysRevLett.98.030405}%
  \BibitemOpen
  \bibfield  {author} {\bibinfo {author} {\bibfnamefont {D.}~\bibnamefont
  {Vitali}}, \bibinfo {author} {\bibfnamefont {S.}~\bibnamefont {Gigan}},
  \bibinfo {author} {\bibfnamefont {A.}~\bibnamefont {Ferreira}}, \bibinfo
  {author} {\bibfnamefont {H.~R.}\ \bibnamefont {B\"ohm}}, \bibinfo {author}
  {\bibfnamefont {P.}~\bibnamefont {Tombesi}}, \bibinfo {author} {\bibfnamefont
  {A.}~\bibnamefont {Guerreiro}}, \bibinfo {author} {\bibfnamefont
  {V.}~\bibnamefont {Vedral}}, \bibinfo {author} {\bibfnamefont
  {A.}~\bibnamefont {Zeilinger}},\ and\ \bibinfo {author} {\bibfnamefont
  {M.}~\bibnamefont {Aspelmeyer}},\ }\bibfield  {title} {\bibinfo {title}
  {Optomechanical entanglement between a movable mirror and a cavity field},\
  }\href {https://doi.org/10.1103/PhysRevLett.98.030405} {\bibfield  {journal}
  {\bibinfo  {journal} {Phys. Rev. Lett.}\ }\textbf {\bibinfo {volume} {98}},\
  \bibinfo {pages} {030405} (\bibinfo {year} {2007})}\BibitemShut {NoStop}%
\bibitem [{\citenamefont {Corbitt}\ \emph {et~al.}(2007)\citenamefont
  {Corbitt}, \citenamefont {Wipf}, \citenamefont {Bodiya}, \citenamefont
  {Ottaway}, \citenamefont {Sigg}, \citenamefont {Smith}, \citenamefont
  {Whitcomb},\ and\ \citenamefont {Mavalvala}}]{PhysRevLett.99.160801}%
  \BibitemOpen
  \bibfield  {author} {\bibinfo {author} {\bibfnamefont {T.}~\bibnamefont
  {Corbitt}}, \bibinfo {author} {\bibfnamefont {C.}~\bibnamefont {Wipf}},
  \bibinfo {author} {\bibfnamefont {T.}~\bibnamefont {Bodiya}}, \bibinfo
  {author} {\bibfnamefont {D.}~\bibnamefont {Ottaway}}, \bibinfo {author}
  {\bibfnamefont {D.}~\bibnamefont {Sigg}}, \bibinfo {author} {\bibfnamefont
  {N.}~\bibnamefont {Smith}}, \bibinfo {author} {\bibfnamefont
  {S.}~\bibnamefont {Whitcomb}},\ and\ \bibinfo {author} {\bibfnamefont
  {N.}~\bibnamefont {Mavalvala}},\ }\bibfield  {title} {\bibinfo {title}
  {Optical dilution and feedback cooling of a gram-scale oscillator to 6.9
  mk},\ }\href {https://doi.org/10.1103/PhysRevLett.99.160801} {\bibfield
  {journal} {\bibinfo  {journal} {Phys. Rev. Lett.}\ }\textbf {\bibinfo
  {volume} {99}},\ \bibinfo {pages} {160801} (\bibinfo {year}
  {2007})}\BibitemShut {NoStop}%
\bibitem [{\citenamefont {Diorico}\ \emph {et~al.}(2022)\citenamefont
  {Diorico}, \citenamefont {Zhutov},\ and\ \citenamefont
  {Hosten}}]{https://doi.org/10.48550/arxiv.2203.04550}%
  \BibitemOpen
  \bibfield  {author} {\bibinfo {author} {\bibfnamefont {F.}~\bibnamefont
  {Diorico}}, \bibinfo {author} {\bibfnamefont {A.}~\bibnamefont {Zhutov}},\
  and\ \bibinfo {author} {\bibfnamefont {O.}~\bibnamefont {Hosten}},\ }\href
  {https://doi.org/10.48550/ARXIV.2203.04550} {\bibinfo {title} {Laser-cavity
  locking at the $10^{-7}$ instability scale utilizing beam elipticity}}
  (\bibinfo {year} {2022})\BibitemShut {NoStop}%
\bibitem [{\citenamefont {Li}\ \emph {et~al.}(2022)\citenamefont {Li},
  \citenamefont {Diorico},\ and\ \citenamefont
  {Hosten}}]{PhysRevApplied.17.054031}%
  \BibitemOpen
  \bibfield  {author} {\bibinfo {author} {\bibfnamefont {V.}~\bibnamefont
  {Li}}, \bibinfo {author} {\bibfnamefont {F.}~\bibnamefont {Diorico}},\ and\
  \bibinfo {author} {\bibfnamefont {O.}~\bibnamefont {Hosten}},\ }\bibfield
  {title} {\bibinfo {title} {Laser frequency-offset locking at 10-hz-level
  instability using hybrid electronic filters},\ }\href
  {https://doi.org/10.1103/PhysRevApplied.17.054031} {\bibfield  {journal}
  {\bibinfo  {journal} {Phys. Rev. Appl.}\ }\textbf {\bibinfo {volume} {17}},\
  \bibinfo {pages} {054031} (\bibinfo {year} {2022})}\BibitemShut {NoStop}%
\bibitem [{\citenamefont {Siegman}(1986)}]{Siegman1986}%
  \BibitemOpen
  \bibfield  {author} {\bibinfo {author} {\bibfnamefont {A.~E.}\ \bibnamefont
  {Siegman}},\ }\href@noop {} {\emph {\bibinfo {title} {Lasers}}}\ (\bibinfo
  {publisher} {University Science Books},\ \bibinfo {year} {1986})\BibitemShut
  {NoStop}%
\bibitem [{\citenamefont {Numata}\ \emph {et~al.}(2003)\citenamefont {Numata},
  \citenamefont {Ando}, \citenamefont {Yamamoto}, \citenamefont {Otsuka},\ and\
  \citenamefont {Tsubono}}]{PhysRevLett.91.260602}%
  \BibitemOpen
  \bibfield  {author} {\bibinfo {author} {\bibfnamefont {K.}~\bibnamefont
  {Numata}}, \bibinfo {author} {\bibfnamefont {M.}~\bibnamefont {Ando}},
  \bibinfo {author} {\bibfnamefont {K.}~\bibnamefont {Yamamoto}}, \bibinfo
  {author} {\bibfnamefont {S.}~\bibnamefont {Otsuka}},\ and\ \bibinfo {author}
  {\bibfnamefont {K.}~\bibnamefont {Tsubono}},\ }\bibfield  {title} {\bibinfo
  {title} {Wide-band direct measurement of thermal fluctuations in an
  interferometer},\ }\href {https://doi.org/10.1103/PhysRevLett.91.260602}
  {\bibfield  {journal} {\bibinfo  {journal} {Phys. Rev. Lett.}\ }\textbf
  {\bibinfo {volume} {91}},\ \bibinfo {pages} {260602} (\bibinfo {year}
  {2003})}\BibitemShut {NoStop}%
\bibitem [{\citenamefont {Gretarsson}\ and\ \citenamefont
  {Harry}(1999)}]{gretarsson1999}%
  \BibitemOpen
  \bibfield  {author} {\bibinfo {author} {\bibfnamefont {A.~M.}\ \bibnamefont
  {Gretarsson}}\ and\ \bibinfo {author} {\bibfnamefont {G.~M.}\ \bibnamefont
  {Harry}},\ }\bibfield  {title} {\bibinfo {title} {Dissipation of mechanical
  energy in fused silica fibers},\ }\href@noop {} {\bibfield  {journal}
  {\bibinfo  {journal} {Review of scientific instruments}\ }\textbf {\bibinfo
  {volume} {70}},\ \bibinfo {pages} {4081} (\bibinfo {year}
  {1999})}\BibitemShut {NoStop}%
\bibitem [{\citenamefont {Harry}\ \emph {et~al.}(2002)\citenamefont {Harry},
  \citenamefont {Gretarsson}, \citenamefont {Saulson}, \citenamefont
  {Kittelberger}, \citenamefont {Penn}, \citenamefont {Startin}, \citenamefont
  {Rowan}, \citenamefont {Fejer}, \citenamefont {Crooks}, \citenamefont
  {Cagnoli}, \citenamefont {Hough},\ and\ \citenamefont
  {Nakagawa}}]{Gregory_M_Harry_2002}%
  \BibitemOpen
  \bibfield  {author} {\bibinfo {author} {\bibfnamefont {G.~M.}\ \bibnamefont
  {Harry}}, \bibinfo {author} {\bibfnamefont {A.~M.}\ \bibnamefont
  {Gretarsson}}, \bibinfo {author} {\bibfnamefont {P.~R.}\ \bibnamefont
  {Saulson}}, \bibinfo {author} {\bibfnamefont {S.~E.}\ \bibnamefont
  {Kittelberger}}, \bibinfo {author} {\bibfnamefont {S.~D.}\ \bibnamefont
  {Penn}}, \bibinfo {author} {\bibfnamefont {W.~J.}\ \bibnamefont {Startin}},
  \bibinfo {author} {\bibfnamefont {S.}~\bibnamefont {Rowan}}, \bibinfo
  {author} {\bibfnamefont {M.~M.}\ \bibnamefont {Fejer}}, \bibinfo {author}
  {\bibfnamefont {D.~R.~M.}\ \bibnamefont {Crooks}}, \bibinfo {author}
  {\bibfnamefont {G.}~\bibnamefont {Cagnoli}}, \bibinfo {author} {\bibfnamefont
  {J.}~\bibnamefont {Hough}},\ and\ \bibinfo {author} {\bibfnamefont
  {N.}~\bibnamefont {Nakagawa}},\ }\bibfield  {title} {\bibinfo {title}
  {Thermal noise in interferometric gravitational wave detectors due to
  dielectric optical coatings},\ }\href
  {https://doi.org/10.1088/0264-9381/19/5/305} {\bibfield  {journal} {\bibinfo
  {journal} {Classical and Quantum Gravity}\ }\textbf {\bibinfo {volume}
  {19}},\ \bibinfo {pages} {897} (\bibinfo {year} {2002})}\BibitemShut
  {NoStop}%
\bibitem [{\citenamefont {Adhikari}(2014)}]{RevModPhys.86.121}%
  \BibitemOpen
  \bibfield  {author} {\bibinfo {author} {\bibfnamefont {R.~X.}\ \bibnamefont
  {Adhikari}},\ }\bibfield  {title} {\bibinfo {title} {Gravitational radiation
  detection with laser interferometry},\ }\href
  {https://doi.org/10.1103/RevModPhys.86.121} {\bibfield  {journal} {\bibinfo
  {journal} {Rev. Mod. Phys.}\ }\textbf {\bibinfo {volume} {86}},\ \bibinfo
  {pages} {121} (\bibinfo {year} {2014})}\BibitemShut {NoStop}%
\bibitem [{\citenamefont {Hosten}\ \emph {et~al.}(2016)\citenamefont {Hosten},
  \citenamefont {Krishnakumar}, \citenamefont {Engelsen},\ and\ \citenamefont
  {Kasevich}}]{hosten2016}%
  \BibitemOpen
  \bibfield  {author} {\bibinfo {author} {\bibfnamefont {O.}~\bibnamefont
  {Hosten}}, \bibinfo {author} {\bibfnamefont {R.}~\bibnamefont
  {Krishnakumar}}, \bibinfo {author} {\bibfnamefont {N.~J.}\ \bibnamefont
  {Engelsen}},\ and\ \bibinfo {author} {\bibfnamefont {M.~A.}\ \bibnamefont
  {Kasevich}},\ }\bibfield  {title} {\bibinfo {title} {Quantum phase
  magnification},\ }\href {https://doi.org/10.1126/science.aaf3397} {\bibfield
  {journal} {\bibinfo  {journal} {Science}\ }\textbf {\bibinfo {volume}
  {352}},\ \bibinfo {pages} {1552} (\bibinfo {year} {2016})},\ \Eprint
  {https://arxiv.org/abs/https://www.science.org/doi/pdf/10.1126/science.aaf3397}
  {https://www.science.org/doi/pdf/10.1126/science.aaf3397} \BibitemShut
  {NoStop}%
\bibitem [{\citenamefont {Nagourney}(2010)}]{nagourney2010}%
  \BibitemOpen
  \bibfield  {author} {\bibinfo {author} {\bibfnamefont {W.}~\bibnamefont
  {Nagourney}},\ }\href@noop {} {\emph {\bibinfo {title} {Quantum electronics
  for atomic physics}}}\ (\bibinfo  {publisher} {OUP Oxford},\ \bibinfo {year}
  {2010})\BibitemShut {NoStop}%
\bibitem [{\citenamefont {Shimoda}\ \emph {et~al.}(2018)\citenamefont
  {Shimoda}, \citenamefont {Aritomi}, \citenamefont {Shoda}, \citenamefont
  {Michimura},\ and\ \citenamefont {Ando}}]{shimoda2018}%
  \BibitemOpen
  \bibfield  {author} {\bibinfo {author} {\bibfnamefont {T.}~\bibnamefont
  {Shimoda}}, \bibinfo {author} {\bibfnamefont {N.}~\bibnamefont {Aritomi}},
  \bibinfo {author} {\bibfnamefont {A.}~\bibnamefont {Shoda}}, \bibinfo
  {author} {\bibfnamefont {Y.}~\bibnamefont {Michimura}},\ and\ \bibinfo
  {author} {\bibfnamefont {M.}~\bibnamefont {Ando}},\ }\bibfield  {title}
  {\bibinfo {title} {Seismic cross-coupling noise in torsion pendulums},\
  }\href@noop {} {\bibfield  {journal} {\bibinfo  {journal} {Physical Review
  D}\ }\textbf {\bibinfo {volume} {97}},\ \bibinfo {pages} {104003} (\bibinfo
  {year} {2018})}\BibitemShut {NoStop}%
\end{thebibliography}%
\end{document}